\documentclass[structabstract]{aa}
\usepackage{graphicx}
\usepackage{txfonts}
\usepackage{natbib}
\usepackage{comment}
\usepackage[usenames,dvipsnames]{color} 
\usepackage{multirow}
\usepackage{enumitem}

\newcommand{\maa}{\alpha \alpha} 
\newcommand{\mga}{\gamma \alpha}
 
\def\Rsolar{$R_{\odot}$}
\def\Msolar{$M_{\odot}$}
\def\Lsolar{$L_{\odot}$}

\newcommand{\perpccubed}{\,{\rm pc^{-3}}}
\newcommand{\perGyr}{\,{\rm Gyr^{-1}}}
\newcommand{\perpcsquared}{\,{\rm pc^{-2}}}
\newcommand{\Mo}{M_{\odot}}
\newcommand{\Ro}{R_{\odot}}

\begin{document}

\title{The binarity of the local white dwarf population }
\titlerunning{The binarity of the local white dwarf population}

\author{S.~Toonen \inst{1,2,3},  M.~Hollands\inst{4}, B.T.~G\"{a}nsicke \inst{4},T.~Boekholt\inst{5,2}}

\institute{
Anton Pannekoek Institute for Astronomy, University of Amsterdam, 1090 GE Amsterdam, The Netherlands
\\ \email{toonen@uva.nl}
\and Leiden Observatory, Leiden University, PO Box 9513, NL-2300 RA Leiden, the Netherlands 
\and Department of Astrophysics, IMAPP, Radboud University Nijmegen, PO Box 9010, 6500 GL Nijmegen, The Netherlands
\and Department of Physics, University of Warwick, Coventry CV4 7AL
\and Departamento de Astronom\'{i}a, Facultad Ciencias F\'{i}sicas y Matem\'{a}ticas, Universidad de Concepci\'{o}n, Av. Esteban Iturra s/n Barrio Universitario, Casilla 160, Concepci\'{o}n, Chile
}

\date{Received 31/10/2016; Accepted 23/02/2017}
\abstract
{As endpoints of stellar evolution, white dwarfs (WDs) are powerful tools to study 
the evolutionary history of the Galaxy. In particular, the multiplicity of WDs contains information regarding the formation and evolution of binary systems.
}
{
Can we understand the multiplicity of the local WD sample from a theoretical point of view? Population synthesis methods are often applied to estimate stellar space densities and event rates, but how well are these estimates calibrated? This can be tested by a comparison with the 20\,pc sample, which contains $\simeq$ 100 stars and is minimally affected by selection biases.
}
{We model the formation and evolution of single stars and binaries within 20\,pc with a population synthesis approach. We construct a model of the current sample of WDs and differentiate between WDs in different configurations, that is single WDs, and resolved and unresolved binaries containing a WD with either a main-sequence (MS) component or with a second WD. We also study the effect of different assumptions concerning the star formation history, binary evolution, and the initial distributions of binary parameters.  
We compile from the literature the available information on the sample of WDs within 20\,pc, with a particular emphasis on their multiplicity, and compare this to the synthetic models.
}
{The observed space densities of single and binary WDs are well reproduced by the models. The space densities of the most common WD systems (single WDs and unresolved WD-MS binaries) are consistent within a factor two with the observed value. We find a discrepancy only for the space density of \emph{resolved} double WDs. 
We exclude that observational selection effects, fast stellar winds, or dynamical interactions with other objects in the Milky Way explain this discrepancy.
We find that either the initial mass ratio distribution in the Solar neighbourhood is biased towards low mass-ratios, or more than ten resolved DWDs have been missed observationally in the 20\,pc sample. 
Furthermore, we show that the low binary fraction of WD systems ($\sim$25\%) compared to Solar-type MS-MS binaries ($\sim$50\%) is consistent with theory, and is mainly caused by mergers in binary systems, and to a lesser degree by WDs hiding in the glare of their companion stars. 
Lastly, Gaia will dramatically increase the size of the volume-limited WD sample, detecting the coolest and oldest WDs out to $\simeq50$\,pc. We provide a detailed estimate of the number of single and binary WDs in the Gaia sample.
}
{}
\maketitle

\section{Introduction}
\label{sec:intro}

As most stars end their life as white dwarfs (WDs), 
they form a significant component of the stellar population and are the most common stellar remnants. As such, 
WD stars play an important role in the study of the structure and the evolutionary history of stellar ensembles \citep{Fon01, Alt10}. 
They provide us with an effective way 
to reconstruct the star formation history (SFH) of the Solar neighbourhood and Galactic disc by analyzing the WD luminosity function \citep[e.g.][]{Tre14}. 
WDs can also be used to constrain with good accuracy the age of 
stellar ensembles, such as the Solar neighbourhood, stellar clusters, and the Galactic disc  \citep{Tor05, Han07, Bed09}. 
Fundamental for these types of studies are observational samples that are as large and homogeneously-selected as possible.

An important, but often complicated aspect in many population studies, is the level of completeness of the observational sample and how to compensate for any observational biases. 
A complete sample of WDs is therefore a powerful tool, but assembling such a sample can be observationally very demanding, as WDs are low-luminosity objects, and the different WD discovery methods, primarily proper motion surveys and ultraviolet excess surveys, have incomplete overlap.
Much time and effort has been devoted to create a complete and volume-limited sample of WDs in the Solar neighbourhood 
\citep[e.g.][]{Hol02, Ven03, kawka04, Far05, kawka06, subasavage07, subasavage08, Hol08, Sio09, Gia12, sayres12, limoges13, limoges15, sion14}. 
The advantage of the Solar neighbourhood is that even the coolest WDs can be identified with relative ease at these short distances from us \citep[e.g.][]{Car14}. The level of completeness that has been achieved for the WD sample within 20\,pc is exceptional, and is estimated to be 80-90\% \citep{Hol08, Sio09, Gia12, holberg16}.

Large and homogeneously-selected samples of stellar systems play a vital role in the empirical verification of population synthesis studies, such as binary population synthesis (BPS)\footnote{See \citet{Too14} for a comparison of four BPS codes.}. 
The BPS approach aims to further improve our understanding of stellar and binary evolution from a statistical point of view, and can aid and further motivate observational surveys. 
It is often used to constrain evolutionary pathways and predict population characteristics, such as event rates or the period distribution
of stellar populations, including type Ia supernovae \citep[for a review see][]{Wan12}, post-common envelope binaries \citep[e.g.][]{Too13, Cam14,Zor14}, or AM CVn  systems \citep[e.g.][]{Nel01c}.
Nonetheless, tests on the number densities of a stellar population (e.g. space densities or event rates) predicted by BPS studies are often not strongly constraining, as the observed number densities are uncertain to (at least) a factor of a few.  However, since the 20\,pc sample of WDs is volume-limited and nearly complete, it allows for a strong test of the number of predicted systems from the BPS method, which is the aim of this paper.

Another important feature of the 20\,pc sample is that it consists of multiple populations of WD systems. It contains WDs formed by single stellar evolution and from mergers in binaries, and WDs in binaries such as double WDs (DWDs) and WD main-sequence binaries (WDMS). 
The sample provides us with a rare opportunity to compare multiple stellar populations, formed from very different evolutionary paths, with the results of self-consistent population synthesis models. 
So far, none of the studies  of the WD luminosity function have included binarity \citep[e.g.][]{Tre14, Tor16}, despite the expected contribution from binaries \citep{Oir14}.

The set-up of this paper is as follows: 
in Sect.~\ref{sec:obs}, we give an overview of the observed sample of local WDs. In Sect.~\ref{sec:bps}, we describe the BPS simulations. 
In Sect.~\ref{sec:res} the self-consistent simulated WD populations are presented. We compare the number of systems in the WD population and its subcomponents 
predicted by the synthetic populations with the observed sample of Sect.~\ref{sec:obs}.
For unresolved binaries, we take into account the selection effects against finding a dim star next to a bright star. We also predict the number of WD systems within 50\,pc in Sect.~\ref{sec:gaia}, which will become available with Gaia. 
In Sect.~\ref{sec:disc} we discuss the hypothesis of missing WD binaries in the Solar neighbourhood, and in Sect.~\ref{sec:concl} our results are summarized.

\section{Observed sample}
\label{sec:obs}

\citet{Hol02} constructed a local WD sample consisting  of 109 WD candidates within 20\,pc. \citet{Hol02} estimated that their sample was approximately 65\% complete. Since then the completeness of the local WD sample was estimated to have risen to 80--90\% \citep{Hol08,Sio09, Gia12}. Most recently, the completeness level has been estimated to be 86\% by \citet{holberg16}.
The local WD sample has been used to derive the local space density $[(4.8 \pm 0.5) \cdot 10^{-3}\,\perpccubed]$ and mass density $[(3.1 \pm 0.3) \cdot 10^{-3}\,\Mo \perpccubed]$ \citep[e.g.][]{Hol02, Hol08, Sio09, holberg16}. 
The kinematical properties of the local WD sample have been studied by \citet{Sio09}, who found that the vast majority of these stars belong to the thin disk. Finally, \citet{Gia12} performed a systematic model atmosphere analysis of all the available data of the local WD population.

 The observed sample that we use here is mainly based on the sample of systems from \cite{Gia12} and full details are given in Appendix\,\ref{sec:app_sample} and Table\,\ref{tbl:obs_single}. The sample of WDs in binaries is given in Table\,\ref{tbl:obs_bin}, and WDs in higher-order systems in Table\,\ref{tbl:obs_ex}. 
A good starting point on WD binarity is provided by 
 \citet{farihi05}, \citet{Hol08}, and \citet{holberg13}. 
We note that the latter paper focuses on Sirius-type binaries (WDs with companions of spectral K and earlier) in the Solar neighbourhood, but is incomplete with respect to low-mass companions. 
 Notes on specific WD systems are given below.

\begin{table*}
\caption{Known WDs in binary systems in the Solar neighbourhood. The distances, spectral types, masses, and luminosities are taken from \citet{Gia12}. References for the binarity of the system are given in the last column. For the unresolved systems, the period $P$ is given in days instead of angular separation, if available. }
\centering
\begin{tabular}{lcccccccc}
\hline
\hline
WD name & Distance [pc] & Spectral  & Mass [\Msolar]  & log L/\Lsolar &Companion &Spectral & Angular & References \\
 &  & type & &  & name& type& separation ["]&   \\
\hline
\multicolumn{9}{c}{\textbf{Resolved WDMS}}  \\
0148+641     & 17.35 (0.15) & DA5.6 & 0.66 (0.03) & -3.08 &  GJ 3117 A &        M2      &12.1 & 1, 2, 3, 4, 5\\
0208$-$510     & 10.782  (0.004) & DA6.9 & 0.59 (0.01) & - &  GJ86A &   K0      &1.9    &  3, 6, 7\\
0415$-$594     & 18.46 (0.05) & DA3.3 & 0.60 (0.02) & - &  eps. Reticulum A       & K2&   12.8     & 7, 9,10\\
0426+588     & 5.51 (0.02) & DC7.1 & 0.69 (0.02) & -3.52 &  GJ 169.1 A       &M4.0   & 9.2   & 4, 5, 8, 11, 12, 13, 14   \\
0628$-$020     & 20.49 (0.46) & DA7.2 & 0.62 (0.01) & - &  LDS 5677B &  M       &         4.5             &  15, 16, 17 \\ 
0642$-$166     & 2.631 (0.009) & DA2.0 & 0.98 (0.03) & -1.53 &  Sirius A        &A0      &7.5    &  8, 10, 18\\
0736+053     & 3.50 (0.01) & DQZ6.5 & 0.63 (0.00) & -3.31 & Procyon A   &F5     & 53      &   10, 19\\
0738$-$172     & 9.096 (0.046) & DZA6.6 & 0.62 (0.02) & -3.35 & GJ 238 B        &M6.5    & 21.4  & 3, 8, 20, 21  \\
0751$-$252     & 17.78 (0.13) & DA9.9 & 0.59 (0.02) & -4.02 &  LTT2976 &M0              &400    &     3, 4, 5, 22, 23\\
1009$-$184     & 18.3 (0.3) & DZ8.3 & 0.59 (0.02) & -3.74 &  LHS 2031 A &K7     &400&        3, 10, 24, 25, 26\\
1043$-$188     & 19.01 (0.18) & DQpec8.7 & 0.53 (0.11) & -3.77 & GJ 401 A        &M3     &       8&      3, 4, 5, 15\\
1105$-$048     &  17.33 (3.75) & DA3.5  & 0.54 (0.01) & - & LP 672-2 & M3 & 279   &    2, 17, 27, 28\\
1132$-$325     & 9.560 (0.034) & DC10 & -  & - &      HD 100623 &K0     &16     &       8, 10, 17, 29\\
1327$-$083     & 16.2 (0.7) & DA3.5 & 0.61 (0.03) & -2.16 &  LHS 353     &M4.5 & 503                     &  3, 28, 30, 31\\
1345+238     & 12.1 (0.3) & DC11.0 & 0.45 (0.02) & -4.08 &   LHS 362    &M5      &199                    &      3, 31, 32\\
1544$-$377     & 15.25 (0.12) & DA4.8 & 0.55 (0.03) & -2.67 &  GJ 599 A   &G6     &       15.2    & 3, 4, 5, 8, 10, 33\\
1620$-$391     & 12.792 (0.062) & DA2.1 & 0.61 (0.02) & -1.12 & HD 147513       &G5      &345&  8, 10, 27, 31, 34\\

1917$-$077     & 10.1 (0.3) & DBQA4.8 & 0.62 (0.02) & -2.81 &  LDS 678B &M6     &       27.3            &  3, 20, 31 \\ 
2011+065     & 22.4 (1.0) & DC7.6 & 0.7 (0.04) & -3.68 & LHS 3533 & M3.5 & 101 &        13, 26, 35\\

2151$-$015     & 24.5 (1.0) & DA5.5 &0.58 (0.03) & -2.96 & LTT 8747B    &M8&            1.082   &   3, 31, 36\\
2154$-$512     & 15.12 (0.12) & DQP8.3 & 0.60 (0.04) & -3.44 & GJ841 A       &M2     & 28.5                  &      3, 4, 5, 30, 37, 38\\
2307+548     & 16.2 (0.7) &  DA8.8 & 0.58 (-) & - & G233-42 &   M5&     6&  13, 17, 39, 40 \\
2307$-$691     & 20.94 (0.38) & DA5 & 0.57 (-) & - & GJ 1280 &  K3 &    13.1&   17 \\
2341+322     & 17.61 (0.55) & DA4.0 & 0.56 (0.03) & -2.3 & G130-6       &M3&     175             &      20, 41, 42\\

\hline
\multicolumn{9}{c}{\textbf{Unresolved WDMS}}  \\
0419$-$487     & 20.13 (0.55) & DA7.8 & 0.22 (0.05) & -3.14 & - &M4 &$P=0.3037$ &   3, 16, 43, 44, 45\\
0454+620     &  21.6 (1.2) &  DA4.6 &  1.14 (0.07) & - & - &-   & -     &   13, 39 \\
\hline
\multicolumn{9}{c}{\textbf{Resolved DWD}}  \\
0648+641 & 33.3 (5.9) & DA8.3 & 0.98 (0.09) & -4.09 &  WD0649+639  & DA8.1 & 490 &  13, 28, 46, this work \\
0747+073A   & 18.3 (0.2) & DC10.4 &0.48 (0.01) & -4.20 & WD0747+073B   & DC12 & 16.4 &  27, 47\\
2126+734     & 21.2 (0.8) & DA3.2 & 0.60 (0.03) & -1.97 & -& DC10 & 1.4 &  13, 31, 48 \\
2226$-$754     & 13.5 (0.9) & DC12.0 & 0.58 (0.00) & -4.32 & WD2226$-$755  & DC12.0 &  93 &                3, 49\\

\hline
\multicolumn{9}{c}{\textbf{Unresolved DWD}}  \\
0135$-$052     & 12.3 (0.4) & DA6.9 & 0.24 (0.01) & -3.00 &   - &DA6.9 &$P=1.56$ &     27, 50\\
0532+414     & 22.4 (1.0) & DA6.5 & 0.52 (0.03) & -3.20 &       - &- &- &              3, 49\\

\multicolumn{9}{c}{\textbf{Unresolved DWD candidate}}  \\
0108+277     & 28.0 (1.5) & DA7.8 &0.59 (0.00) & -3.60 & -& -& -&                   3\\ 
0121$-$429     & 18.3 (0.3) & DAH8.0 &0.41 (0.01) & -3.46 & -& -& -&                 3\\
0423+120     & 17.4 (0.8) & DC8.2 &0.65 (0.04) & -3.75 & -& -&- &                  3,25\\
0503$-$174     & 21.9 (1.9) & DAH9.5 &0.38 (0.07) & -3.75 & -& -& -&                  3\\
0839$-$327     & 8.80 (0.15) & DA5.6 &0.44 (0.07) & -2.84 & -& -& -&                 3, 8\\
2048+263     & 20.1 (1.4) & DA9.9 &0.24 (0.04) & -3.65 &- &- & -&                   3\\
2248+293     & 20.9 (1.9) & DA9.0 &0.35 (0.07) & -3.62 & -& -& -&                    3\\
2322+137     & 22.3 (1.0) & DA9.7 &0.35 (0.03) & -3.75 &- &- & -&                   3 \\
\hline
\end{tabular} 
\newline
\begin{flushleft} \tablefoot{
$^1$\citet{greenstein70};
$^2$\citet{wegner81}; 
$^3$\citet{Gia12};
$^4$\cite{Tre17};
$^5$\cite{Gai16};
$^6$\citet{mugrauer05};
$^{7}$\citet{vanleeuwen07};
$^{8}$ http://www.DenseProject.com;
$^9$\citet{farihi11b};
$^10$\citet{holberg13};
$^{11}$\citet{liebert76};   
$^{12}$\citet{heintz90};
$^{13}$\citet{limoges15};
$^{14}$\citet{dieterich12};
$^{15}$\citet{oswalt88};
$^{16}$\citet{subasavage09};
$^{17}$\citet{holberg16};
$^{18}$\citet{gatewood78};
$^{19}$\citet{liebert13};
$^{20}$\cite{luyten49};         
$^{21}$\citet{davison15};
$^{22}$\citet{subasavage08};
$^{23}$\citet{luyten80};
$^{24}$\citet{henry02};  
$^{25}$\citet{Hol08};
$^{26}$\citet{hawley96};
$^{27}$\citet{sion14};
$^{28}$\citet{vanaltenabook95};
$^{29}$\citet{poveda94};
$^{30}$\citet{eggen56};
$^{31}$\citet{farihi05};
$^{32}$\citet{dahn76};
$^{33}$\citet{wegner73};
$^{34}$\citet{alexander69};
$^{35}$\citet{giclas59};
$^{36}$\citet{farihi06};
$^{37}$\citet{vornanen10};
$^{38}$\citet{tamazian14};
$^{39}$\citet{limoges13};
$^{40}$\citet{newton14};
$^{41}$\citet{sion88};
$^{42}$\citet{garces11};
$^{43}$\citet{bessell79};
$^{44}$\citet{bruch98};
$^{45}$\citet{maxted07};
$^{46}$\citet{lepine05};
$^{47}$\citet{greenstein70};
$^{48}$\citet{zuckerman97};
$^{49}$\citet{scholz02};
$^{50}$\cite{saffer98}
$^{51}$\citet{zuckerman03}.
} \end{flushleft}
\label{tbl:obs_bin}
\end{table*}

\subsection{Notes on individual objects}
\label{sec:notes}

\subsubsection{A new resolved double degenerate at 33 pc}
\label{sec:disc_0649}
We report the identification of a new resolved double degenerate system, comprising WD0648+641 and the recently discovered WD0649+639. The two stars are  8.2 arcmin apart and their proper motions are (432, -142) mas/yr  and (421, -130) mas/yr, respectively \citep{lepine05}. 
The trigonometric distance to WD0648+641 has been determined to be $33\pm5$\,pc \citep{vanaltenabook95}, and the spectroscopic distance to WD0649+639 is about 21\,pc \citep{limoges13, limoges15}. 
Nevertheless, since the temperatures, spectroscopic masses,
and V-band magnitudes of both WDs are very comparable
 \citep[$6220\pm137$K versus $6050\pm 98$K, $0.87\pm0.15$\Msolar versus $0.98\pm 0.09$\Msolar, and 14.67 versus 15.07 for WD0649+639 and WD0648+641, respectively, see][]{limoges15}, we deem it likely that the two WDs are at a comparable distance.

\subsubsection{Distances}
\label{sec:distance}
The distances given in Table\,\ref{tbl:obs_bin} are based on \citet{Gia12} with updates from \citet{limoges13}, \citet{limoges15}, and the Discovery and Evaluation of Nearby Stellar Embers (DENSE) project\footnote{http://www.DenseProject.com}. For a few systems, the derived distances from different studies are significantly discrepant, such that their membership of the 20\,pc sample is ambiguous. We discuss these systems here in detail.  

\begin{itemize}
\item WD0019+423 has a spectroscopic distance of $12.9\pm3.0$\,pc \citep{limoges15}.
However, its $V$-band magnitude of 16.5, effective temperature of 5590 K, and log $g$ of 8.0 from \citet{limoges15} implies an absolute magnitude of 14.5 (using the WD models as described in Sect.\,\ref{sec:mag}) and a distance of 25\,pc. This system is therefore removed from the 20\,pc sample.

\item WD0454$+$620 is an unresolved WDMS system in which the M-dwarf contaminates the WD spectrum. Both \cite{limoges13} and \cite{limoges15} take special care in the fitting procedure of the WD spectral lines, however, the derived distances are distinct. The most recent measurement of \cite{limoges15} gives a distance of 
$21.6 \pm1.2$\,pc, which gives a 10\% chance for the system to be within 20\,pc. With the  distance found by \cite{limoges13} ($24.9\pm0.9$\,pc) it is excluded that WD0454$+$620 is within 20\,pc. We adopt the most recent value of \cite{limoges15}, however, we note that this does not significantly affect our conclusions of Sect.\,\ref{sec:concl}.

\item WD1242$-$105 has recently been shown not to be a single object, but to be part of a double degenerate binary \citep{debes15} with a short period of 2.85\,hr. 
These authors find a trigonometric distance of $39 \pm 1$\,pc, which excludes  WD1242$-$105 from the 20\,pc sample. Previously, the distance to  WD1242$-$105 was estimated to be $23.5 \pm 1$\,pc \citep{Gia12}, based on spectral model fitting assuming a single object.

\item  Regarding WD1657+321, \citet{Gia12} find a distance of $>$50\,pc when assuming a log $g$ of 8.0. On the other hand, \citet{kawka06} derive  log $g = 8.76\pm0.20$ and a distance $d=22$\,pc. 
\citet{kawka06} do not provide an uncertainty on the distance. We tentatively assume an uncertainty of $\pm$1\,pc, which gives a 3\% probability for WD1657+321 to be within 20\,pc.
Even with an uncertainty of 2\,pc on the distance estimate of \citet{kawka06} (and subsequently a probability of 20\% of being a member of the 20\,pc sample,), the space density within 20\,pc does not change in a significant way. 

\item  For WD1912+143, we adopt the trigonometric distance $35 \pm 6.6$\,pc \citep{dahn1982, limoges15}, which effectively excludes it from the 20\,pc sample. This value is in agreement with the trigonometric distance found by \citet{vanaltenabook95} of $36.2 \pm 7.5$\,pc, significantly exceeding the spectroscopic distance found by \citet{limoges13} of $19.4 \pm 0.7$\,pc.

\item WD2011+065 has a trigonometric distance of 
$22.4\pm 1.0$\,pc  based on the parallax measurement of $44.7 \pm 1.9$ mas \citep{vanaltenabook95, bergeron97}. Notably, \citet{limoges15} find a larger uncertainty on the distance (2.4\,pc) based on the same parallax measurement.
In the former case, there is a $\sim$1\% chance that WD2011+065 falls within 20\,pc, whereas an uncertainty of 2.4\,pc gives a chance of about 15\%. In both cases, WD2011+065 does not significantly contribute to the space density within 20\,pc.

\item WD2151$-$015 is part of a binary with a MS companion \citep{farihi05, farihi06, Hol08}. The binary has  been resolved with an angular separation of 1.082$\pm$0.002" \citep{farihi06}. The distance found by \citet{Gia12} of 24.5$\pm$1.0\,pc places it well outside 20\,pc, however, other estimates place it on the boundary of the 20\,pc sample, for example 21\,pc by \citet{farihi06} and 20.97$\pm$1.21\,pc by \citet{Hol08}. The latter gives a 20\% probability for the system to be within 20\,pc.
\end{itemize}

\subsubsection{Double WD candidates}
\label{sec:dwd}
A number of systems are classified as (unresolved) DWD candidates in Table\,\ref{tbl:obs_bin}. These are:
\begin{itemize}
\item WD0423+120 which is overly bright for its parallax \citep{Hol08} and therefore considered to be a DWD candidate by these authors. Both the parallax and photometric distances (17.36\,pc vs 11.88\,pc, respectively), position the system within 20\,pc from the Sun. 
\item WD0839$-$327 which is classified as a DWD candidate due to possible radial variations in the DA star \citep{bragaglia90}. This claim is supported by the marginal difference in the photometric and trigonometric distance (7\,pc and  $8.87 \pm 0.77$\,pc respectively) found by \citet{kawka07}. The trigonometric distance as given by DENSE is $8.80 \pm 0.15$\,pc (see Tbl.\ref{tbl:obs_bin}). \citet{holberg08b} found a photometric distance of $8.05 \pm 0.11$\,pc. 
\item WD2048+263 which is suspected to be a double-degenerate system by \citet{bergeron01} based on the low-gravity and mass, as well as the suspected dilution of the Balmer H$\alpha$ profile of the
visible DA WD by a possible DC companion.  
\item WD0108+277, WD0121$-$429, WD0839$-$327, WD0503$-$174, WD2054$-$050, and WD2248+293 which are suggested to be double degenerates by \citet{Gia12}. This is based on the low mass they derive by means of the photometric technique. 
The masses are too low for stars to have evolved as single stars ($\lesssim 0.5$\,\Msolar). 
For the same reason we add WD2322+137, however, it has a low probability of being within 20\,pc (i.e. 1\%). If WD2054$-$050 is indeed a DWD, then the system would be a triple system with an MS companion in a wide orbit \citep{greenstein86a, greenstein86b, sion88, Hol08}. 
\item WD0322$-$019 has been considered a close DWD in the past, however, \citet{farihi11} showed that the source of line broadening was magnetism and not binarity.
\end{itemize}

A word of caution is necessary for the mass estimates of WDs in unresolved binaries (and candidates). The mass estimates in Tables\,\ref{tbl:obs_bin}~and~\ref{tbl:obs_ex} are taken from \citet{Gia12}, who fitted single WD models to all spectra in the 20\,pc sample. For example, \citet{Gia12} note that the spectrum of WD0419$-$487 (RR Caeli) is contaminated by the presence of an M-dwarf companion. As a consequence the WD mass according to \citet{Gia12} is significantly lower ($0.22\pm0.05\,\Mo$) than that found by \citet{maxted07} ($0.440\pm0.023\,\Mo$). \citet{maxted07} determined the mass and radius of WD0419$-$487 from the combined analysis of the radial velocities and the eclipse light curve.

\subsubsection{Questionable multiplicity}

For eleven WDs, it has been suggested that they are part of a binary or multiple system, however,  
confirmation or follow-up is lacking. In more detail:
\begin{itemize}
\item WD0148+467 is listed as WD+MS in \citet{Hol08} based on the Hipparcos \& Tycho catalogues. We are unable to find any other objects in these catalogues within two  degrees that have a similar parallax and proper motion to WD0148+467. 
\item WD0310$-$688 is suggested to have a second component in the Washington Double Star catalogue. \citet{stauffer10} suggest the companion does not exist.
\item \citet{probst83} found a possible common proper motion companion for WD0341+182, that is BPM31492. 
\item \citet{hoard07} report a tentative low mass companion for WD0357+081.  
\item WD0426+588 is in a wide binary (Stein2051) with an M-star companion. There is some suggestion that this is a triple system 
\citep{strand77}. In their model, the red component is an astrometric binary. 
\item WD0644+375 is a single WD now, but \citet{ouyed11} speculate it used to be a neutron star-WD binary, where the neutron star transitioned to a quark star during a quark nova, enriching the WD with iron, and stripping some of the WD mass. If this is the case, it should be excluded from the comparison with the BPS models, as in these models the evolution of neutron stars is not taken into account. 
\item WD0856+331 was previously identified as being part of a common proper motion binary with HD77408 \citep{wegner81}. However, the magnitudes of the proper motions \citep{lepine05} and the parallaxes \citep{vanaltenabook95,vanleeuwen07} differ significantly.  
\item WD1142$-$645 is listed by \citet{Hol08} as a binary, however, we do not find this to be supported by the associated references or any other literature. 
\item WD1647+591 shows possible radial velocity variability for this system \citep{saffer98}, however, as the parallax and photometric distance agree to within 1.2 sigma \citep{vanleeuwen07, Hol08}, we consider it a single WD. 
\item There is some confusion in the literature as to the multiplicity of the system containing WD1917$-$077. At the time of writing, SIMBAD lists this as a quadruple system. The supposed D component appears in the Washington Double Star catalogue, however its proper motion differs significantly from the others. The star listed as the C component appears in various literature \citep{turon93, gould04, lampens07} where it is found to have the same proper motion as the A/B component. However, the B/C components were at the time spatially very close leading to blending, which may have impacted their analyses. Comparison of images between DSS1 and DSS2 surveys show only the A/B components to have any detectable motion between the two epochs laying to rest any suggestion of higher multiplicity. 
\item \citet{saffer98} found WD2117+539 to have possible RV variability, however \citet{foss91} did not find variability. 
\end{itemize}

\subsubsection{Triples and quadruples}
\label{sec:ex}

\begin{table*}
\caption{Known WDs in the Solar neighbourhood that are part of triples and quadruples. The distances, spectral types, masses, and luminosities are taken from \citet{Gia12}. For the unresolved systems, the period $P$ is given instead of angular separation. }
\centering
\begin{tabular}{lcccccccc}
\hline
\hline
 & Distance [pc]& Spectral &  Mass [\Msolar] &  log L/\Lsolar & Companion  & Spectral & Angular & References \\

 & &  type &   &   &  name &  type &  separation ["]& \\

\hline
0101+048     & 21.3 (1.7) & DA6.3 & 0.36 (0.05) & -2.96 & 
- & DC & see text & 1, 2, 3, 4\\
&&&&&HD 6101&K3+K8&1276& \\
0326$-$273     & 17.4 (4.3) & DA5.9 & 0.45 (0.18) & -2.97 & 
- & DC8 & $P=1.88d$ & 4, 5, 6 \\
&&&&&GB 1060B&M3.5&7&  \\
0413$-$077     & 4.984 (0.006) & DA3.1 & 0.59 (0.03) & -1.85 & 
40 Eri A & K0.5 & 83.4 & 7, 8, 9, 10 \\
&&&&& 40 Eri C & M4.5 & 11.9 & \\
0433+270     & 17.48 (0.13) & DA9 & 0.62 (0.02) & -3.87 & V833 Tau &       K2$^{14}$& 123.9                &  4, 8, 11, 12, 13, 14\\
0727+482A    & 11.1 (0.1) & DC10 & 0.51 (0.01) & -4.01 & 
WD0727+482B & DC10.1 & 0.656 &  9, 15, 16, 17  \\
&&&&& G107-69 & M5$^{*}$ & 103.2 & \\
0743$-$336     & 15.2 (0.1) & DC10.6 & 0.55 (0.01) & -4.23 & 171 Pup A  &F9      &870&    8, 18, 19, 20 \\
1633+572     & 14.4 (0.5) & DQpec8.1 & 0.57 (0.04) & -3.75 & 
CM draconis& M4.5$^{*}$ & 26 & 20, 21 \\
2054$-$050     & 16.09 (0.14) & DC11.6 & 0.37 (0.06) & -4.11 & 
Ross 193 & M3.0 & 15.1 & 4, 10, 11, 12, 22, 23 \\ 
2351$-$335     & 22.90 (0.75)$^{9,25}$ & DA5.7 & 0.58 (0.03) & -3.03 & 
LDS826B & M3.5 & 6.6 & 4, 6, 24, 25, 26  \\ 
&&&&& LDS826C & M8.5 & 103 &\\
\hline
\end{tabular} 
\begin{flushleft} \tablefoot{
$^1$\cite{saffer98}
$^2$\cite{maxted00};
$^3$\cite{caballero09};
$^{4}$\cite{Gia12};
$^5$\cite{nelemans05};
$^6$\cite{luyten49};
$^{7}$\cite{holberg12};
$^{8}$\cite{holberg13};
$^{9}$\cite{sion14};
$^{10}$Discovery and Evaluation of Nearby Stellar Embers (DENSE) project, http://www.DenseProject.com;
$^{11}$\cite{Tre17}; 
$^{12}$\cite{Gai16};
$^{13}$\cite{hartmann81};
$^{14}$\cite{tokovinin06};
$^{15}$\cite{strand76};
$^{16}$\cite{harrington81};
$^{17}$\cite{buscombe98};
$^{18}$\cite{hartkopf12};
$^{19}$\cite{tokovinin12b}; 
$^{20}$\cite{limoges15};
$^{21}$\cite{morales09};
$^{22}$\cite{vanbiesbroeck61};
$^{23}$\cite{tamazian14};
$^{24}$\cite{scholz04};
$^{25}$\cite{farihi05};
$^{26}$\citet{subasavage09};
$^{*}$Spectral type corresponds to an unresolved binary;
\label{tbl:obs_ex}
}\end{flushleft}
\end{table*}

There are a few WDs found in triples and quadruples (Table\,\ref{tbl:obs_ex}). 
 The structure of observed multiples tend to be hierarchical, for example triples consist of an inner binary and a distant companion star \citep{Hut83}.  
Despite the distance between the companion and the binary, the evolution of these systems can be different from that of isolated binary systems \citep{Too16}. For example, \citet{Tho11} shows that the dynamical effect of a third companion on compact DWD binaries can lead to an enhanced rate of mergers and type Ia supernovae. The BPS models presented in this paper do not include the possible interaction of a distant companion. 
For completeness, we discuss WDs in multiples separately from isolated WDs and binaries 
 in the comparison between the synthetic and observed populations in Sect.\,\ref{sec:res_num}. 
Because there are only $\sim$6 WDs in multiples within 20\,pc, including or excluding these systems does not significantly change our conclusions.

 The high-order systems are the following:
\begin{itemize}
\item WD0101+048 is part of a hierarchical quadruple, consisting of a close DWD binary \citep{maxted00} and an MS-MS binary \citep{caballero09}. 
The double MS-binary is a visual binary with a period of $\sim$29 yr and an angular separation of $\sim$0.5 mas \citep{balega06}. There is some uncertainty regarding the period of the close DWD, however a period of 1.2\,d or 6.4\,d is most likely \citep{maxted00}.   
\item WD0326$-$273 is a close DWD \citep{zuckerman03, nelemans05} with an M 5 star in a wide orbit \citep{sion88, poveda94, garces11}. 
\item WD0413$-$077 is part of a resolved WDMS binary, with a K-star companion in a wide orbit \citep{wegner88, tokovinin08}.
\item WD0433$+$270 is the outer companion of a spectroscopic binary of spectral type K2 \citep{tokovinin06, zhao11, holberg13}. 
The K-binary may also have a planetary mass companion at 0.025" separation \citep{lucas02, holberg13}.
\item WD0727$+$482 is in a quadruple system. This system consists of a resolved DWD, and an unresolved MS-MS binary of spectral type M \citep{harrington81,probst83, sion91, andrews12, janson14}.
\item WD0743$-$336 is the outer star in a triple system \citep{tokovinin12}. 
The inner system, 171 Pup, is an astrometric binary and is resolved with 
speckle interferometry. 
\item WD1633$+$572 is in a wide orbit around an eclipsing MS-MS binary of spectral type M \citep{silvestri02, sion88, poveda94, feiden14}.
\item For WD2054$-$050, see Sect.~\ref{sec:dwd}.
\item WD2351$-$335 is part of a triple system \citep{scholz04, farihi05}. The inner binary is a visual pair consisting of the WD and an M 3.5-star with a separation of 6.6". The outer star is a M 8.5 star in a wide orbit of about 100". 
\end{itemize}

\subsubsection{Miscellaneous}
WD0939+071 is not included in our sample, because it was mistakenly  classified as a WD \citep{gianninas11, Gia12}. The star is also known as GR 431 and PG 0939+072 and is reclassified by \citet{gianninas11} to be an MS F-type star. WD0806$-$661 is included as a single star ignoring its brown dwarf companion \citep{luhman11}.

\section{Stellar and binary population synthesis}
\label{sec:bps}

\subsection{SeBa  - a fast stellar and binary evolution code}
\label{sec:seba}

We employ the population synthesis code SeBa \citep{Por96, Nel01, Too12, Too13} to simulate a large number of single stars and binaries. We use SeBa to evolve stars from the zero-age main sequence (ZAMS) until and including the remnant phase. At every timestep, processes such as stellar winds, mass transfer, angular momentum loss, common envelope, magnetic braking, and gravitational radiation are considered with appropriate recipes. SeBa is incorporated into the Astrophysics Multipurpose Software
Environment, or AMUSE. This is a component library with a homogeneous
interface structure and can be downloaded for free at {\tt
amusecode.org} \citep{Por09}.

In this paper, we employ 12 BPS models. The BPS models are the 2x2x3 possible permutations of two models for the SFH (BP \& cSFR), two models for the initial period distribution (`Abt' \& `Lognormal'), and three models for the common-envelope phase ($\mga$, $\maa,$ \& $\maa$2). These assumptions affect the predicted space densities most compared to other uncertainties regarding the evolution and formation of stars and binaries. The models are explained in detail in the following sections and an overview is given in Table\,\ref{tbl:models}.

\begin{table*}
\caption{Overview of different BPS models. There are two models for the SFH, two for the period distribution, and three for the CE-phase, giving 12 models in total.}
\begin{tabular}{llcc}
\hline
\hline
 & Model &Description &  Reference\\
\hline
\multirow{3}{*}{Star formation history} &\multirow{2}{*}{BP} & Star formation rate and space density depends on time and location   & 1 \\
 & & in the Galaxy. SFR peaks at early times, declines afterwards &  \\
&cSFR & Constant space density and SFR for 10 Gyr & -\\
\hline
\multirow{3}{*}{Initial period distribution} 
&Abt & Log-uniform  & 2\\
&Lognormal & Lognormal distribution with a mean of 5.03\,d & 3 \\
\hline
\multirow{3}{*}{Common-envelope phase }&$\mga$ & $\gamma=1.75$, $\alpha\lambda= 2$; Preferred for unresolved DWDs& 4,5,6\\
&$\maa$ & $\alpha\lambda= 2$ & 4,6 \\
&$\maa$2& $\alpha\lambda= 0.25$; Preferred for unresolved WDMS& 7,8,9\\
\hline
\end{tabular}
\label{tbl:models}
\begin{flushleft}
\tablefoot{
$^{1}$\citet{Boi99}; 
$^{2}$\citet{Abt83}; 
$^{3}$\citet{Rag10};
$^{4}$\citet{Nel00};
$^{5}$\citet{Nel01};
$^{6}$\citet{Too12};
$^{7}$\citet{Zor10};
$^{8}$\citet{Too13};
$^{9}$\citet{Cam14}. 
}\end{flushleft}
\end{table*}

\subsection{The initial stellar population}
\label{sec:init_pop}
The initial stellar population is generated on a Monte Carlo based approach, according to appropriate distribution functions. 
The initial mass of single stars and of binary primaries are drawn between 0.95--10\,\Msolar~from a Kroupa initial mass function (IMF) \citep{Kro93}. 
Furthermore, Solar metallicities are assumed.
For binaries, unless specified otherwise, the secondary mass is drawn from a uniform mass ratio distribution between 0 and 1 \citep{Duc13}, and the eccentricity
from a thermal distribution \citep{Heg75} between 0 and 1. 
For the orbital period (or equivalently the semi-major axis) distribution, we adopt two models. For model `Abt', the orbits are drawn from a power-law distribution with an exponent of $-1$ \citep{Abt83} ranging from 0 to 10$^6$\,\Rsolar. For model `Lognormal', periods are drawn from a lognormal distribution with a mean of 5.03\,days, a dispersion of 2.28 \citep{Rag10}, and a maximum period of $10^{10}$d. 
For Solar-type stars, the latter distribution has become the preferred distribution \citep{Duq91,Rag10,Duc13,Tok14}.

\subsection{Initial binary fraction}
\label{sec:bin_fr}

Observational studies have shown that the binary fraction depends on the spectral type of the primary star \citep[e.g.][]{Sha02, Rag10, Duc13}. Due to the properties of the IMF and SFH, the average WD progenitor is a $\sim$2\,\Msolar~(A-type) star for the WD systems under consideration in this paper. 

For G- and F-type stars observed binary fractions are $44\pm2\%$ \citep{Duc13} and $54\pm2\%$ \citep[][more specifically $50\pm4\%$ for F6--G2 stars and $41\pm3\%$ for G2--K3 stars]{Rag10}. Studies of OB-associations have shown binary fractions of over $70\%$ for O- and B-type stars \citep{Sha02, Kob07, Kou07, San12}. 
From the most thorough search for companions to A-stars \citep{derosa2014}, a binary fraction of $43.6 \pm 5.3$\% is estimated.

In this paper, we assume an initial binary fraction of 50\% unless specified otherwise. 
If an initial binary fraction $f$ other than 0.5 is shown to be appropriate, the predicted number of systems (see Table\,\ref{tbl:numbers_wds}) 
can easily be adjusted as follows: the number of binaries and merged systems should be multiplied with the correction factor $w_{\rm bin}$, and the number of single WDs with $w_{\rm sin}$. The correction factors are given by:
\begin{equation}
w_{\rm sin}=\frac{\langle M_{\rm sin}\rangle + \langle M_{\rm bin}\rangle}{\langle M_{\rm sin}\rangle +\langle M_{\rm bin}\rangle f/(1-f)},
\label{eq:bf_ss}
\end{equation}
and 
\begin{equation}
w_{\rm bin} = \frac{\langle M_{\rm sin}\rangle +\langle M_{\rm bin}\rangle }{ \langle M_{\rm sin}\rangle (1-f)/f + \langle M_{\rm bin}\rangle},
\label{eq:bf_bin}
\end{equation}
 where $\langle M_{\rm sin}\rangle$ is the average mass of a single star and $\langle M_{\rm bin}\rangle $ the average (total) mass of a binary system. Assuming the initial distributions as described in Sect.\,\ref{sec:init_pop} and the full range in stellar masses of 0.1--100\,\Msolar, $\langle M_{\rm sin}\rangle =0.49\,\Mo$~and $\langle M_{\rm bin}\rangle =0.74\,\Mo$ for the period distribution of \citet{Abt83}, and $\langle M_{\rm sin}\rangle =0.52\,\Mo$~and $\langle M_{\rm bin}\rangle =0.78\,\Mo$ for the lognormal period distribution.
 
For a lower limit on the binary fraction of $40\%$, the correction factors are $w_{\rm bin} =0.83 $ and $w_{\rm sin} = 1.25$ for both period distributions. For an upper limit of $60\%$, the correction factors are $w_{\rm bin} =1.15 $ and $w_{\rm sin} = 0.77$.
The uncertainty in the initial binary fraction therefore induces an error on the BPS results of about 15--25\%

\subsection{Common-envelope evolution}
\label{sec:ce}
An important phase in the evolution of many binary systems occurs when one or both stars fill their Roche lobes, and matter can flow from the donor star through the first Lagrangian point to the companion star. As the evolutionary timescales are shorter for more massive stars, the most massive component of the binary will reach the giant phase first, and is likely to fill its Roche lobe before the companion does. If the mass transfer rate from the donor star increases upon mass loss, a runaway situation ensues, named the common-envelope (CE) phase \citep{Pac76}. The CE-phase is a short-lived phase in which the envelope of the donor star engulfs the companion star. If sufficient energy and angular momentum is transferred to the envelope, it can be expelled, and a merger of the binary can be avoided. 
The CE-phase plays an essential role in binary star evolution, in particular, in the formation of short-period systems. 
The orbital outcome is one of the aspects of binary evolution that affects the synthetic binary populations most \citep[e.g.][]{Too13}. Despite its importance and the enormous efforts of the community, the CE-phase is not understood in detail.

The classical model for the CE-phase is the $\alpha$-formalism, which is based on the energy budget \citep{Tut79}. The $\alpha$-parameter describes the efficiency with which orbital energy is consumed to unbind the CE according to
\begin{equation}
E_{\rm gr} = \alpha (E_{\rm orb,init}-E_{\rm orb,final}),
\label{eq:alpha-ce}
\end{equation}
where $E_{\rm orb}$ is the orbital energy and $E_{\rm gr}$ is the binding energy of the envelope.  The orbital and binding energy are as defined in \citet{Web84}, where $E_{\rm gr}$ is approximated by
\begin{equation}
E_{\rm gr} = \frac{GM_{\rm d} M_{\rm d,env}}{\lambda R},
\label{eq:Egr}
\end{equation} 
with $M_{\rm d}$ the donor mass, $M_{\rm d, env}$ the envelope mass of the donor star, $\lambda$ the envelope-structure parameter, and $R$ the radius of the donor star. Due to the uncertainty in the value of both $\alpha$ and $\lambda$, they are often combined into one parameter $\alpha\lambda$. 

An alternative method for CE-evolution, is the $\gamma$-formalism \citep{Nel00}, which is based on angular momentum balance. The $\gamma$-parameter describes the efficiency with which orbital angular momentum is used to expel the CE according to
\begin{equation}
\frac{J_{\rm b, init}-J_{\rm b,  final}}{J_{\rm b,init}} = \gamma \frac{\Delta M_{\rm d}}{M_{\rm d}+ M_{\rm a}},
\end{equation} 
where $J_{\rm b,init}$ and $J_{\rm b,final}$ are the orbital angular momentum of the pre- and post-mass transfer binary respectively, and $M_{\rm a}$ is the mass of the companion. 
The motivation for the $\gamma$-formalism comes from the observed mass-ratio distribution of DWD systems that could not be explained by the $\alpha$-formalism nor stable mass transfer for a Hertzsprung gap donor star \citep[see][]{Nel00}. The idea is that angular momentum can be used for the expulsion of the envelope when there is a large amount of angular momentum available, such as in binaries with similar-mass objects. However, the physical mechanism remains unclear. 
Interestingly, \citet[][see also \citealt{Woo10_Myk}]{Woo12} suggested an alternative model to create double WDs. This evolutionary path involves stable, non-conservative mass transfer between a red giant and an MS star. The effect on the orbit is a modest widening with a result alike to the $\gamma$-description. Further studies have to take place to see if this path suffices to create a significant number of DWDs.

In this paper, we adopt three distinct binary evolution models 
that differ in their treatment of the CE-phase. The models are based on different combinations of the $\alpha$- and $\gamma$-formalism with different values of $\alpha\lambda$ and $\gamma$ (see Table\,\ref{tbl:models}). In detail:
\begin{itemize}
\item
In model $\alpha\alpha$, the $\alpha$-formalism is used to determine the outcome of every CE-phase. 
The value of the $\alpha\lambda$-parameter ($\alpha\lambda = 2$) is based on \citet{Nel00}, who deduced this value from reconstructing the second phase of mass transfer for observed DWDs. 
\item
For model $\mga$, the $\gamma$-prescription is applied unless the 
binary contains a compact object
or the CE is triggered by a tidal instability rather than dynamically unstable Roche lobe overflow \citep[see][]{Too12}. The value of the $\alpha\lambda$-parameter is equal to that in model $\maa$. 
The value of the $\gamma$-parameter  ($\gamma = 1.75$) is based on modelling the first phase of mass transfer of observed DWDs \citep{Nel00}. 
\item Model $\maa2$ is similar to model $\maa$, but with a low value of $\alpha\lambda$ ($\alpha\lambda = 0.25$), such that the binary orbit shrinks more strongly during the CE-phase. 
The motivation for model $\maa$2 comes from the population of close WDMS, that is post-common envelope binaries. 
With various techniques \citet{Zor10}, \citet{Too13}, and \citet{Cam14} have shown that the common-envelope phase proceeds less efficiently than is typically assumed in these systems, implying a smaller value for $\alpha\lambda$. This finding is based on the concentration of the observed period-distribution at short periods ranging from a few hours to a few days, but a lack of systems at longer periods \citep[e.g.][]{Neb11}.

\end{itemize}

\subsection{Star formation history}
\label{sec:gal}

    \begin{figure}
    \centering
        \includegraphics[width=\columnwidth, clip=true, trim =0mm 0mm 0mm 135mm]{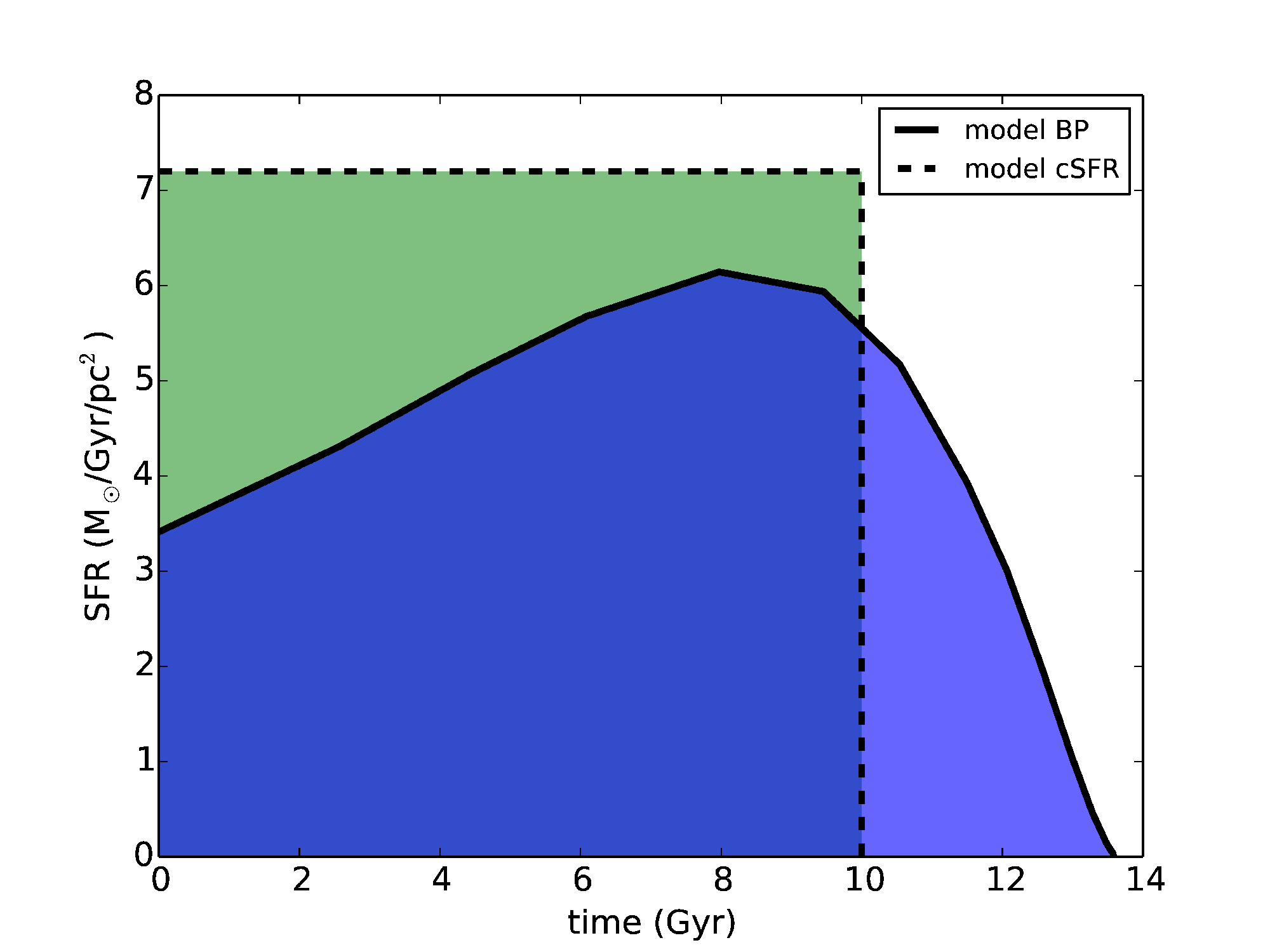} 
    \caption{Star formation rate as a function of time for model BP and model cSFR. Regarding model BP, the star formation rate at a Galactocentric distance of 8.5 kpc is shown. To convert the local star formation rate of model cSFR to \Msolar\,$\perGyr \perpcsquared$, a Galactic scale height of 300\,pc is assumed \citep{Roe07b, Roe07a}. }
    \label{fig:sfr}
    \end{figure}
    
Regarding the assumptions about the Galaxy, two models are adopted that differ in their treatment of the SFH. This comprises the formation rate of the stars and their assigned positions in the Milky Way. 

Model BP is taken from \citet[][based on \citealt{Nel04}]{Too13}. In this model the star formation rate is a function of time and position in the Galaxy \citep{Boi99}. 
It peaks early in the history of the Galaxy and has decreased substantially since then. 
We assume the Galactic scale height of our binary systems to be 300\,pc \citep{Roe07a, Roe07b}. The Galactic star formation rate as a function of time (averaged over space) is shown in the left panel of Fig.\,2 in \citet{Nel04}. For this project, only the star formation rate in the Solar neighbourhood is relevant which is shown in Fig.\,\ref{fig:sfr}. It peaks around 8\,Gyr, and extends to  13.5\,Gyr, which \citet{Boi99} assume is the age of the Galactic disk. However, from MS and WD populations, it has been shown that oldest stars within the disk have an age of  8-10\,Gyr \citep[e.g.][]{Osw96, bergeron97, Del05, Sal09, haywood13, gianninas15}.

Model cSFR is a more simplistic model of the Milky Way with a constant star formation rate and a homogeneous spatial distribution of stellar systems in the Solar neighbourhood. The star formation rate is normalized, such that the total stellar mass in the Galaxy (in the full mass range of 0.1--100\,\Msolar) is $6\cdot 10^{10}$\,\Msolar. 
The spatial distribution is normalized in such way that a spherical region of radius $x$ centred on the Sun contains a fraction of systems in the Galaxy equal to $(4\pi x^3)/( 3V)$, where $V$ is the Galactic volume of $5 \cdot 10^{11}$\,pc$^3$. 
We note that from a more elaborate model distribution of stars in the Galaxy, which is dependent on the Galactocentric distance, \citet{Nel01} found a similar relation between the local space density and the total number of stars in the Galaxy (their Eq.3), that is, $V=4.8\cdot 10^{11}$\,pc$^3$.
For model cSFR, we assume star formation has proceeded for the last 10 Gyr. 
This time span is appropriate for the thin disk, where the majority of objects in the 20\,pc sample  are located \citep{sion14}. 
The average star formation rate (SFR) in mode cSFR is 6\,\Msolar~yr$^{-1}$(see also Fig.\,\ref{fig:sfr}).

\subsection{Magnitudes}
\label{sec:mag}
The absolute magnitudes (bolometric, as well as $ugriz$-bands) are taken from the WD cooling curves of pure hydrogen atmosphere models \citep[][and references therein]{Hol06, Kow06, Tre11}. 
For MS stars we adopt the absolute $ugriz$-magnitudes as given by \citet{Kra07}.
For both the MS stars and WDs, we linearly interpolate between the brightness models. For those stars that are not included in the grids of brightness models, the closest gridpoint is taken. $V$-band magnitudes are calculated as a transformation from the $g$- and $r$-magnitude according to \citet{Jes05} for stars.

\subsection{Types of white dwarf systems}
\label{sec:types}
In this paper we consider six types of stellar systems containing WDs:
\begin{itemize}
\item Single star: A star that begins and ends its life as a single star.\\
\item Merger: A single WD that has formed as a result of a merger in a binary system. \\
\item Resolved WDMS: A binary consisting of a WD and a main-sequence (MS) component in a wide orbit. We assume an orbit can be resolved if the angular separation is larger than the critical angular separation $s_{\rm crit}$:
\begin{equation}
\mathrm{log}(s_{\rm crit}) = 0.04556|\Delta V|-0.0416,
\label{eq:ang_sep_crit}
\end{equation}
where $\Delta V$ is the difference in the $V$-band magnitude of the two stellar components of the binary and $s_{\rm crit}$ in arcseconds. The critical angular separation is an empirical limit that takes into account the brightness contrast between the stars. It is a fit through the three most compact, resolved binaries (Fig.\,\ref{fig:logSep_dv}) in our sample of WDMS and DWDs within 20\,pc. For our standard model we exclude the multiple system WD0727+482 at 0.656", as this system is only marginally resolved \citep{strand76}. For our optimistic and pessimistic scenario of resolving binaries, we translate the critical separation to 
\begin{equation}
\mathrm{log}(s_{\rm crit, opt}) = 0.04556|\Delta V|-0.1968,
\label{eq:ang_sep_crit_pos}
\end{equation}
such that a binary similar to WD0727+482 would just be resolved in our data, and

\begin{equation}
\mathrm{log}(s_{\rm crit, pes}) = 0.04556|\Delta V|+0.3010,
\label{eq:ang_sep_crit_neg}
\end{equation}
such that a binary with $\Delta V=0$ is resolved only if the angular separation exceeds 2". \\

    \begin{figure}
    \centering
        \includegraphics[width=\columnwidth, clip=true, trim =0mm 0mm 0mm 0mm]{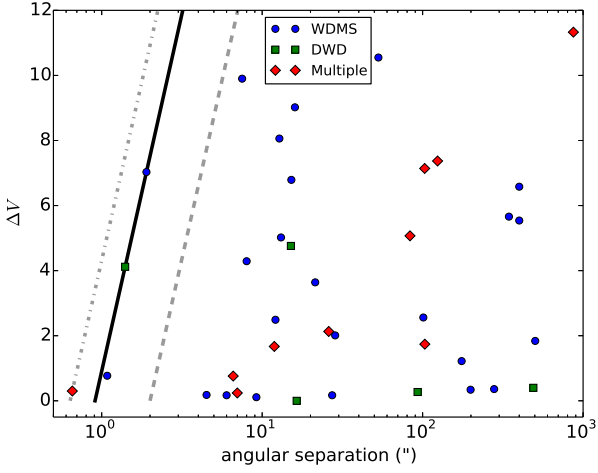} 
    \caption{$V$-band magnitude difference as a function of angular separation for the resolved orbits of WDs in Tables\,\ref{tbl:obs_bin}~and~\ref{tbl:obs_ex}. 
    Resolved WDMS are shown with blue circles and DWDs with green squares. The resolved orbits in triples and quadruples are shown with red diamonds. The resolved orbits in multiples mainly consist of a WD with an MS-companion (see Table\ref{tbl:obs_ex}).
    Overplotted are our empirical estimates of the  critical angular separation $s_{\rm crit}$. Our standard model of Eq.\,\ref{eq:ang_sep_crit} is shown as the black solid line, our optimistic model as the grey dashed-dotted line (Eq.\,\ref{eq:ang_sep_crit_pos}), and our pessimistic model as the grey dashed line (Eq.\,\ref{eq:ang_sep_crit_neg}). }
    \label{fig:logSep_dv}
    \end{figure}

\item Unresolved WDMS: A binary consisting of a WD and an MS in an orbit with an angular separation less than $s_{\rm crit}$. 

This population contains binaries that have undergone a phase of mass transfer (such as post-common-envelope binaries) as well as systems in which no mass transfer has taken place. 
The observed sample of WDMS is strongly affected by selection effects. We assume that unresolved WDMS can only be observed as a WDMS when both components are visible, that is, when 

\begin{equation}
\Delta g \equiv g_{\rm WD}-g_{\rm MS}<1,
\label{eq:g_bias}
\end{equation}
and 
\begin{equation}
\Delta z \equiv z_{\rm WD}-z_{\rm MS} > -1, 
\label{eq:z_bias}
\end{equation}
where $g$ and $z$ represent the magnitudes in the Sloan \textit{g}- and \textit{z}-bands of the WD and MS component. We note that in this paper the term 'unresolved WDMS' refers to an unresolved WDMS in which both components are visible, unless stated differently. 
\\

\item Resolved DWD: A binary consisting of two WDs in an orbit with an angular separation larger than $s_{\rm crit}$. These binaries are all sufficiently wide such that mass transfer does not take place at any point in their evolution.\\
\item Unresolved DWD: A binary consisting of  two WDs in an orbit with an angular separation less than $s_{\rm crit}$. We assume an unresolved DWD can be distinguished from a single WD if both stars contribute significantly to the light, that is, when 
\begin{equation}
\Delta r \equiv |r_{\rm WD1}-r_{\rm WD2}| < 1,
\label{eq:r_bias}
\end{equation}
where $r$ represents the magnitudes in the Sloan \textit{r}-band of each of the WD components (WD component 1 and 2). As for unresolved WDMS, the term 'unresolved DWDs' is used in this paper for those unresolved DWDs where both components contribute to the light, unless stated differently. 
\end{itemize} 

Other types of WD binaries are not taken into account in this project, such as  binaries that are currently interacting (e.g. cataclysmic variables or AM CVn systems) or binaries with evolved stars, neutron stars, or black holes as companions. These systems have not been observed in the Solar neighbourhood, and it is likely that they are much less numerous in general than the binaries considered in this paper. 

For the synthetic binaries, the angular separation $s$ on the sky is calculated according to
\begin{equation}
s=\frac{a(1+e^2/2)}{2d},
\label{eq:ang_sep}
\end{equation}
where $a$ is the semi-major axis, $e$ is the eccentricity of the orbit, and $d$ the distance from us to the binary given by the Galactic model (Sect.\,\ref{sec:gal}). The time-averaged distance between the two stars for a given orbit is $a(1+e^2/2)$. The factor two arises from averaging over all possible orientations on the sky.

\section{White dwarfs within 20\,pc} 
\label{sec:res}

\label{sec:res_num}

\begin{table*}
\caption{Number of systems with WDs components within 20\,pc, see also Fig.\,\ref{fig:nr}. 
The observed sample is based on \cite{Gia12}, but see Sect.~\ref{sec:obs} for adaptations. For unresolved DWDs, we list two numbers. The first number represents confirmed DWD systems, whereas the number in brackets represents the number of confirmed plus candidate DWDs. The third line lists the number of WD systems in triples and quadruples, which are not included in the first line. The evolution of these systems has not been simulated in the BPS models. 
The different BPS models are described in Sect.\,\ref{sec:bps} and an overview is given in Table\,\ref{tbl:models}. The selection effects described in Sect.\,\ref{sec:types} have been applied to the BPS models. Single WDs are formed by single stellar evolution and mergers in binaries. As such, for a given BPS model, the sum of the `Single stars' column and the `Mergers' column should be compared with the observed number of single WDs. 
The statistical errors on the BPS simulations are given in brackets. }
\centering
\begin{tabular}{lll|cccccc}
\hline
\hline
\multicolumn{9}{c}{\textbf{Observations}}  \\
\hline
&& & \multicolumn{2}{c}{Single WDs}  & \multicolumn{2}{c}{WDMS}  & \multicolumn{2}{c}{DWD}  \\
 &  &  &   \multicolumn{2}{c}{ }   & Resolved  & Unresolved & Resolved  & Unresolved \\
\hline
\textbf{Observed} & &  & \multicolumn{2}{|c}{$96.5\pm3.0$} & $19.2\pm 0.4$ &$0.5\pm0.6$ & $2.1\pm0.3$& $1.0\pm 0.1$ ($ 5.0\pm0.8$)  \\
86\% complete & &  & \multicolumn{2}{|c}{$112$} & $22$ &$0.58$ & $2.4$& $1.2$ ($5.8$)  \\
In multiples & &  & \multicolumn{2}{|c}{-} &  $4.0\pm 0.01$& 0 & $1.0\pm0.0$& $1.0\pm 0.6$ ($2.0 \pm 0.6$ )\\
\hline
\hline
\multicolumn{9}{c}{\textbf{BPS models} }  \\
\hline
 &  &  & Single WDs & Mergers  & \multicolumn{2}{c}{WDMS}  & \multicolumn{2}{c}{DWD}  \\
SFH & Period distr. & CE  &   &   & Resolved  & Unresolved & Resolved  & Unresolved \\
\hline
\multirow{3}{*}{BP} & \multirow{3}{*}{Abt} & $\mga$ & \multirow{3}{*}{126 (3.5)} &36 (1.9)&\multirow{3}{*}{30 (0.8)}&2.4 (0.21)&\multirow{3}{*}{20 (0.63)}&8.2 (0.40)\\
 &  & $\maa$ & &43 (2.1) &&2.3 (0.21) &&4.0 (0.28)\\
 &  & $\maa2$ & &50 (2.2)&&1.3 (0.16)&&2.0 (0.20)\\
\hline
\multirow{3}{*}{BP} & \multirow{3}{*}{Lognormal} & $\mga$ &\multirow{3}{*}{126 (3.5)} &15 (1.2)&\multirow{3}{*}{40 (0.9)}&2.5 (0.22)&\multirow{3}{*}{28 (0.75)}&8.0 (0.40) \\
 &  & $\maa$ & &19 (1.4)&&2.4 (0.22)&&4.0 (0.28)\\
 &  & $\maa2$ & &28 (1.7)&&1.5 (0.17)&& 2.3 (0.22)\\
\hline \hline
\multirow{3}{*}{cSFR} & \multirow{3}{*}{Abt} & $\mga$ & \multirow{3}{*}{89 (0.5)}&26 (0.1)&\multirow{3}{*}{22 (0.23)}&1.8 (0.07)&\multirow{3}{*}{15 (0.06)}& 6.1 (0.04)\\
 &  & $\maa$ & &30 (0.1) &&1.9 (0.07)&&3.1 (0.03)\\
 &  & $\maa2$ & &38 (0.1) &&1.0 (0.05)&& 1.5 (0.02)\\
\hline
\multirow{3}{*}{cSFR} & \multirow{3}{*}{Lognormal} & $\mga$ &\multirow{3}{*}{89 (0.5)} &12 (0.05)&\multirow{3}{*}{29 (0.27)}&1.9 (0.07) &\multirow{3}{*}{21 (0.07)}& 5.8 (0.04)\\
 &  & $\maa$ & &14 (0.06) &&2.0 (0.07)&&3.0 (0.03)\\
 &  & $\maa2$ & &21 (0.07) &&1.2 (0.05)&& 1.7 (0.02)\\
\hline
\end{tabular}
\\
\label{tbl:numbers_wds}
\end{table*}

Table\,\ref{tbl:numbers_wds} shows the number of WD systems within 20\,pc as predicted by the BPS approach for different models of the Galaxy, different initial period distributions, and different models of common-envelope evolution. 
The error on the synthetic number of WD systems in Table\,\ref{tbl:numbers_wds} represents the statistical error in the simulations. It is estimated by the square root of the total number of systems of that stellar type in the simulations. 
We have simulated multiple realisations of the local WD populations, which reduces the statistical errors of the BPS models. 
Besides statistical errors, systematic errors originate due to the uncertainties in binary formation and evolution. The systematic errors dominate over the statistical errors in our simulations. For this reason, statistical errors are often omitted in BPS studies; instead different models of binary evolution are compared to gain insight into the systematic errors. 

In Table\,\ref{tbl:numbers_wds}, we show the effect of different CE-models, but only for merger systems, unresolved WDMS, and unresolved DWDs; as single stars, resolved WDMS and DWDs are not affected by binary evolutionary processes.
The most common systems are purely single stars, followed by mergers (in a binary leading to a single WD) and resolved WDMS.  
The predicted population of resolved WDMS is larger than the population of resolved DWDs, because not all stars will become a WD within a Hubble time. 
On the other hand, the predicted population of unresolved WDMS is smaller than the population of unresolved DWDs. This is because the observational selection effects on WDMS are much stronger than in DWDs (see Sect.\,\ref{sec:types}). 
In our simulations, 8--19 unresolved WDMS (1 in $\sim$1.15)\footnote{There are three candidates for these systems which have been detected based on astrometric perturbations of M-dwarfs \citep{Del99, winters16} within 20\,pc. The WD companions have not been detected photometrically so far.} and 0.5--2 unresolved DWDs are discarded (1 in 4-5.5) because of the selection effects of Eqs.\,\ref{eq:g_bias}-\ref{eq:z_bias}. 
Only very few unresolved DWDs are discarded, which means that the WD components of these DWDs tend to have relatively similar brightnesses. We find that this is because the sample is volume-limited instead of magnitude-limited.

For each type of WD system, the observed number of systems within 20\,pc is shown in Table\,\ref{tbl:numbers_wds}. This table also gives a first-order correction for the incompleteness of the 20\,pc sample, based on the completeness estimate of \citet{holberg16} of 86\%.  
Table\,\ref{tbl:numbers_wds} also lists the number of WD binaries that are part of triples and quadruples.

The observed number of systems within 20\,pc is based on Tables\,\ref{tbl:obs_bin}, \ref{tbl:obs_ex},~and~\ref{tbl:obs_single}. For each system, we calculate the probability that the system is within 20\,pc with a Monte Carlo approach that takes into account the uncertainty in the distance as given by column 3 of Tables\,\ref{tbl:obs_bin}, \ref{tbl:obs_ex},~and~\ref{tbl:obs_single}.
As a consequence, some systems with a mean distance just outside of 20\,pc have a non-zero probability of being within 20\,pc. 
And equally, some systems inside, but close to, the 20 pc boundary have a non-zero chance to fall outside our sample.
The number of systems within 20\,pc is then estimated by the sum of the probability of each system. The errors on the number of systems within 20\,pc are based on the same Monte Carlo study. These errors do not include any uncertainty regarding the binarity of the known systems, that is, whether any of the single WDs have an unseen companion or not. Furthermore, these errors do not take into account the uncertainty due to low number statistics.

\subsection{Single white dwarfs}

\begin{figure*}
    \centering
    \begin{tabular}{cc}
        \includegraphics[width=0.4\textwidth,clip=true, trim =0mm 0mm -15mm 10mm]{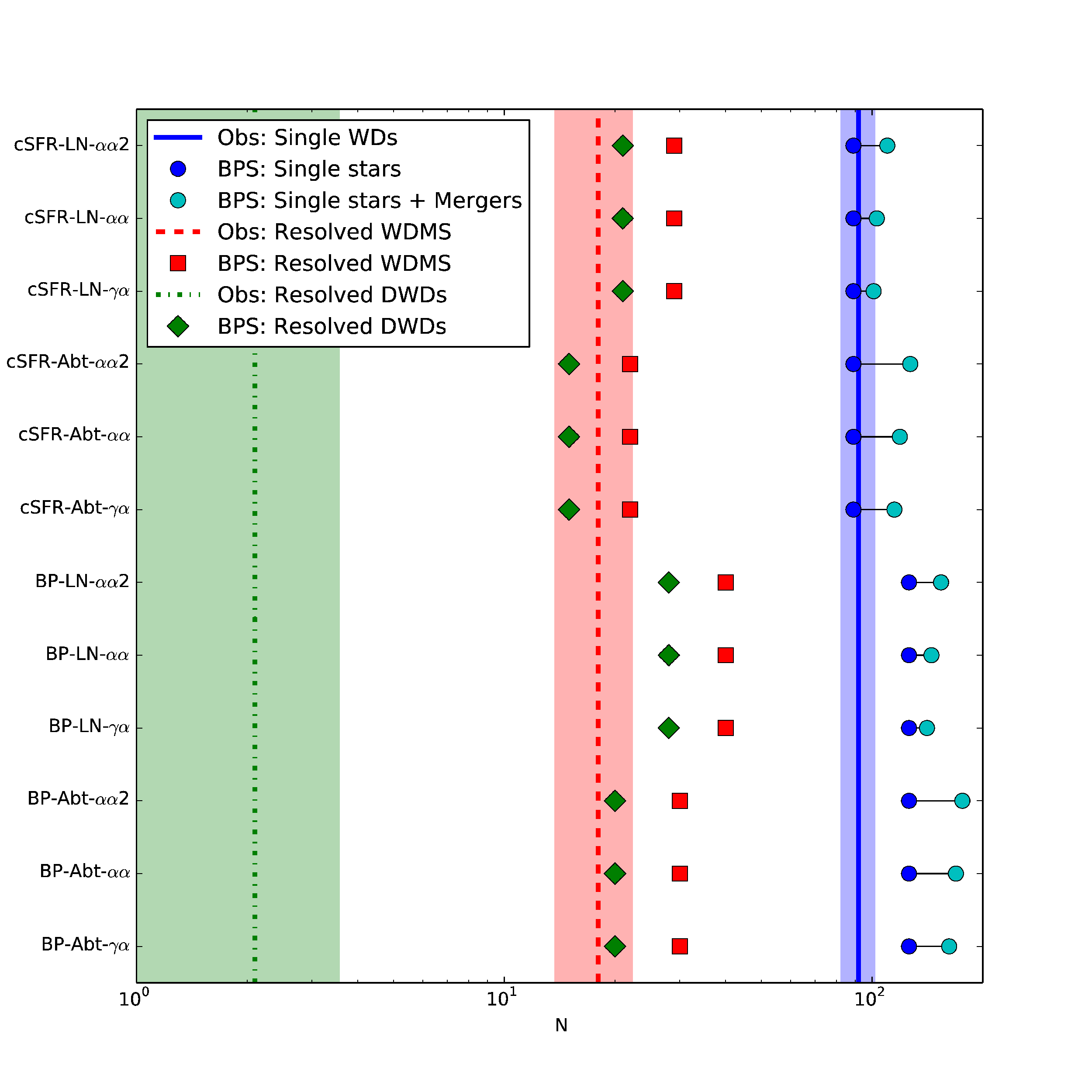} &
        \includegraphics[width=0.4\textwidth, clip=true, trim =0mm 0mm -15mm 10mm]{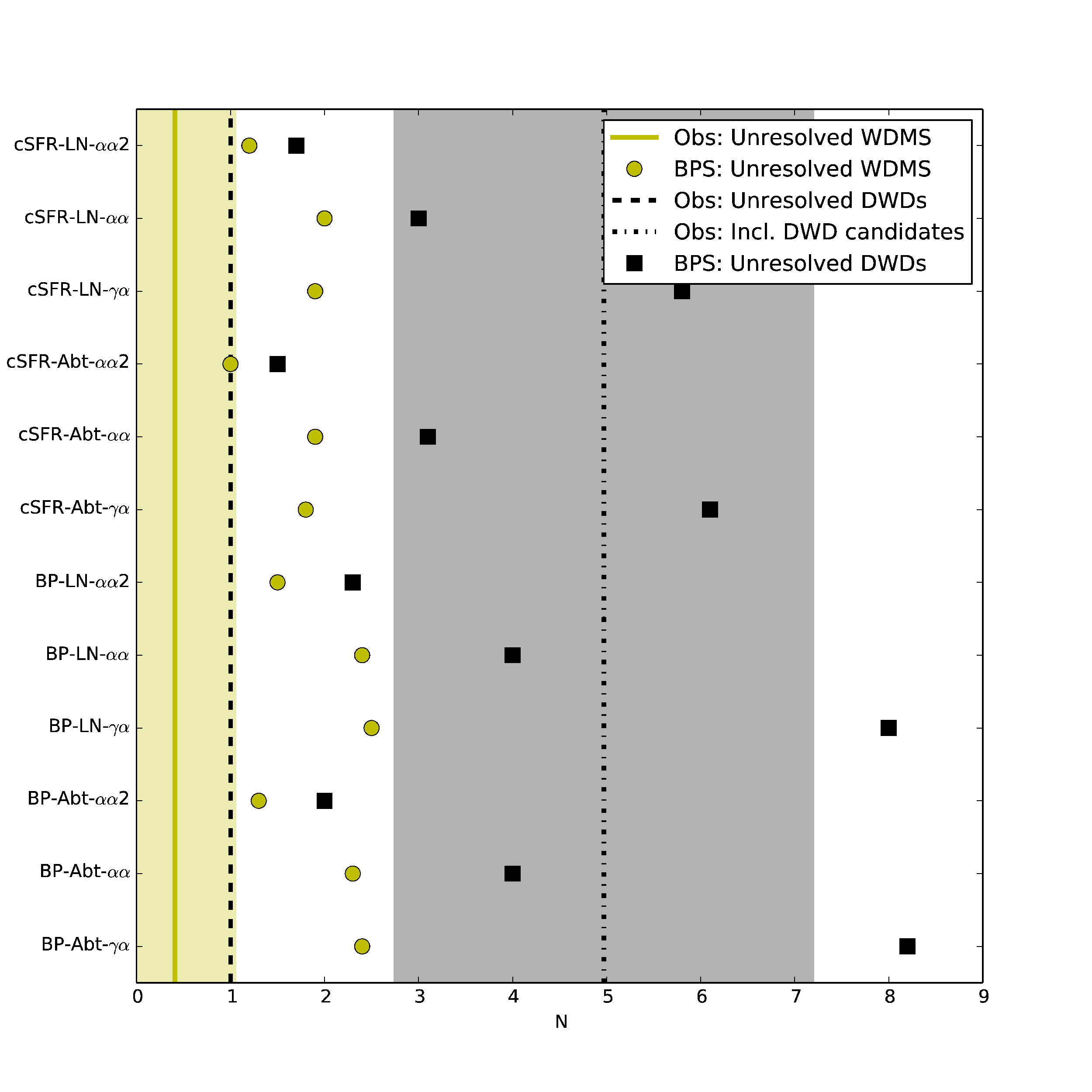} \\
        (a) & (b)\\
        \end{tabular}
    \caption{Comparison of the known number of WD systems with that of the synthetic models. On the left, the comparisons for single WDs and resolved binaries are shown, on the right for unresolved binaries. 
    The lines represent the observations and the markers the BPS models. The shaded area around the lines represents the statistical error on the observations from the square-root law. The statistical error is larger than the error given in Table\,\ref{tbl:numbers_wds} based on the distance estimate of individual systems. }
    \label{fig:nr}
    \end{figure*}

Single WDs mostly descend from isolated single stars, but can also be formed from binaries in which the stellar components merge. 
Comparing the observations with the combination of the two channels (Fig.\,\ref{fig:nr}a), our models predict roughly the same number of WDs (within a factor of 1.8, i.e. 96.1 and 101--176, respectively). Taking into account an 86\% completeness level of the observed sample, this factor reduces to 1.6.

The fraction of single WDs from mergers is not insignificant (10--30\% of all single WDs). This is consistent with estimates for the halo \citep{Oir14}. 
Additionally, this evolutionary channel is interesting in the context of magnetic WDs. 
A recent hypothesis for strong magnetic fields in single WDs considers a magnetic dynamo generation during a CE-merger in a binary \citep{Tou08}. 
The fraction of magnetic WDs amongst all WDs is poorly estimated due to selection effects, but it ranges from  $21\pm 8$\% within 13\,pc and $13\pm 4$\% within 20\,pc from \citet{kawka07}, to 8\% from \citet{sion14}. This is consistent with the incidence of mergers in our models, but see \citet{Bri15} for a more detailed study.

The synthetic number of single WDs is sensitive to the input assumptions of our models. 
The different models for the SFH affect the predicted number of single WDs (excluding mergers) by a factor of 1.4. 
The number of merged systems is most dependent on the initial distribution of periods, and to a lesser degree on the physics of the CE-phase. Regarding the former, in the adopted log-normal distribution, fewer binaries are formed with (relatively short) periods that result in mergers as compared to model `Abt'. Regarding the latter, when the CE-phase leads to a stronger shrinkage (which increases from model $\mga$, to $\maa$, to $\maa$2), the CE-phase is more likely to lead to a merger of the stellar components.

\subsection{Unresolved WDMS}
The selection effects of unresolved WDMS systems affects the population strongly; only in about 1 of 1--8 systems are both components visible. As a result, our population models predict 1.0--2.5 unresolved WDMS systems to be visible within 20\,pc. The different models for the initial period distribution of the binaries and SFH hardly affect the number of unresolved WDMS.

Our modelling of the selection effects introduces a systematic uncertainty in the synthetic population of WDMS (see Eqs.\,\ref{eq:g_bias}~and~\ref{eq:z_bias}). Equation 9 distinguishes WDMS from apparent single MS; equation 10 distinguishes WDMS from apparent single WDs.\footnote{In most systems the light of the binary is dominated by that of the MS star, and therefore we ignore those WDMS that appear as single WDs in the comparison with the observed sample.} Neither varying the cut between $\Delta z >0$ and $\Delta z >-2$, nor making a cut in the \textit{i}-band instead of the \textit{z}-band significantly affects the number of unresolved WDMS.
Varying the cut between $\Delta g <0$ and $\Delta g < 2$ leads to a decrease of systems by about 25--42\% and an increase by about 40--63\%, respectively. This is in good agreement with the results of \citet{Too14}.

The boundaries that we apply to differentiate between resolved and unresolved binaries (Eqs.\,\ref{eq:ang_sep_crit}-\ref{eq:ang_sep_crit_neg}) do not affect the number of predicted unresolved WDMS significantly. 
In the optimistic scenario of Eq.\,\ref{eq:ang_sep_crit_pos}, where binaries can be resolved to smaller angular separations then in the standard scenario of Eq.\,\ref{eq:ang_sep_crit}, the number of unresolved WDMS decreases by 7--13\%. In the pessimistic scenario in which binaries can be resolved only down to an angular separation of 2" (Eq.\,\ref{eq:ang_sep_crit_neg}), the number of unresolved WDMS decreases by 14--31\%.

For compact WDMS that have gone through a CE-phase (i.e. post-common envelope binaries or PCEBs), the preferred CE-model is $\maa2$ (Sec.\,\ref{sec:ce}).
From these models, 1.0--1.5 WDMS systems are predicted within 20\,pc, and 0.7--2.5 including the uncertainty in selection effects. This is consistent with the observed number of $0.5\pm0.6$ from Table\,\ref{tbl:numbers_wds} (see also Fig.\,\ref{fig:nr}b). The number is based on one unresolved WDMS (WD0419$-$487 or RR Caeli) that is on the edge of 20\,pc with $d=20.13 \pm 0.55$\,pc.

Without the distance restriction of the 20\,pc sample, the observed lower limit on the space density of PCEBs is $(6-30)\cdot 10^{-6}\,\perpccubed$ \citep{Sch03}. 
In our models the space density of visible, unresolved WDMS with $P<100$\,d (i.e. PCEBs) is $(4.0-16)\cdot 10^{-6}\,\perpccubed$.
These space densities are calculated in a cylindrical volume with height
above the plane of 200\,pc and radii of 200\,pc and 500\,pc centred
on the Sun. We require both stars to contribute to the light according to Eqs.\,\ref{eq:g_bias}~and~\ref{eq:z_bias}, and the WDMS to be brighter than 20th magnitude in the $g$-band. 
Furthermore, the space density is only calculated for the BPS models that are based on the SFH of \citet{Boi99} (model BP\footnote{For model BP the space density of systems goes down when one averages over a larger volume (further away from the plane of the Galaxy).}), as the homogeneous spatial distribution of stars assumed in model cSFR is not valid at large distances from the Galactic plane. 
In \citet{Too13} the space density of visible PCEBs was simulated using some of the same models as in this paper, that is, based on the SFH of \citet{Boi99} (model BP) and the initial period distribution from \citet{Abt83} (model `Abt').
 Depending on which volume is averaged over, and whether model $\mga$, $\maa$~or $\maa$2 is applied for the CE-phase,  the space density that \citet{Too13} find ranges between $(4.0-15)\cdot 10^{-6}\,\perpccubed$. Both theoretical space densities are in good agreement with the observed space density of PCEBs.

\subsection{Unresolved DWDs}
The models presented in this paper predict  $\simeq$1.5--8 unresolved DWDs within 20\,pc. In the 20\,pc sample, there is only one confirmed (isolated) unresolved DWD, WD0135$-$052, which is in agreement with the lower limit of predicted DWD numbers based on model $\maa$2 (Fig.\,\ref{fig:nr}b). Including those WDs that have been classified as DWD candidates (Sect.\,\ref{sec:obs}) increases the observed number to  $5\pm1$, in good agreement with our models. 
Besides these DWD candidates, there are five systems (WD0141$-$675, WD1223$-$659, WD1632+177, WD2008$-$600, and WD2140+207) whose masses are very close to the lower limit from single stellar evolution ($\lesssim 0.5$\,\Msolar), which might have an undetected companion. 
Additionally, there are two confirmed DWDs (WD0101+048 and WD 0326$-$273), that are part of higher-order systems, and one DWD candidate with an MS companion (WD2054$-$050). Given the large uncertainty in the total number of unresolved DWDs, it is not possible to place a strong constraint on the BPS models. We can only conclude that the models are consistent with the observed numbers within the uncertainties.

The different models for the SFH or initial period distribution of the binaries hardly affect the number of unresolved DWDs. The major uncertainty is the CE-phase with the three different models varying by about a factor of 3--4. 
The preferred model of CE-evolution for DWDs is model $\mga$ (Sect.\,\ref{sec:ce}), which predicts the highest number of DWDs. 
Varying the boundary between resolved and unresolved DWD affects the number of systems by less than a factor 2. For the optimistic scenario of resolving binaries, the number of unresolved DWDs decreases by 10--30\% depending on the CE-model. For the pessimistic scenario, the number increases by 16--24\% for model $\mga$, 35--46\% for model $\maa$, and most strongly for model $\maa$2 with an increase of 73--84\%.

The effect of the uncertainty in the theoretical selection effects applied to the synthetic population of unresolved DWDs (Eq.\,\ref{eq:r_bias}) is small.
Varying the cut $\Delta r$ between 0.5, 1.5, and 2 compared to the standard of 1, leads to a decrease of 20--43\%, an increase of 8--14\%, and an increase of 13--20\%, respectively. Overall, the majority of close DWDs satisfy the $r$-magnitude criterion of Eq.\,\ref{eq:r_bias} in the BPS models. In other words, 
in most cases both WDs contribute to the light and only a few systems are discarded from the synthetic models. 
Depending on the model, 0.5--2.3 systems (18--30\%) are removed from the synthetic models to satisfy Eq.\,\ref{eq:r_bias}. 
Including these systems as an apparent single WD does not change the number of single WDs significantly. Therefore we refrain from adding these systems to the apparently single WDs in the comparison in this paper.

When lifting the distance restriction of 20\,pc, \citet{Max99} find a 95\% probability that the fraction of double degenerates among DA WDs lies in the range 1.7--19\%. Based on the ESO Supernova Type Ia Progenitor surveY (SPY) survey, the fraction of unresolved DWDs compared to all WDs is $7 \pm 1$\%  (priv. comm. Tom Marsh). 
Additionally, the binary fraction of DWDs has been measured from a statistical method \citep{Mao12} by measuring the maximum radial velocity shift between observations of the same WD. From the Sloan digital sky survey (SDSS), a binary fraction of 3-20\% has been derived for separations less
than 0.05AU \citep{Bad12}, and for the SPY survey a fraction of 10.3\% $\pm$ 1.7\% (random uncertainty) $\pm$ 1.5\% (systematic uncertainty) for separations less than 4AU \citep{Mao16}.
Assuming a fraction of 5--10\% holds for the Solar neighbourhood, one would expect five--ten close DWDs in the 20\,pc sample. The number of unresolved DWDs could even be higher as the radial velocity studies of \citet{Max99} and SPY are not sensitive to the full range of periods in our unresolved DWD category ($\lesssim 50-100$d). 
In summary, a number of five--ten close DWDs is in good agreement with our models, in particular the preferred CE-model for DWDs, model $\mga$. Furthermore, it might indicate that some of the DWD candidates are indeed DWDs.

\subsection{Resolved WDMS and DWD binaries}
\label{sec:res_num_resolved}
The predicted number of resolved WDMS and DWDs ranges from about 20--40 and 15--30. The uncertainties on the predicted space densities from the synthetic models are about a factor of $\simeq$ 2. This uncertainty comes from the different models used for the SFH and initial period distribution. 
The effect of varying the boundary between resolved and unresolved binaries affects the number of resolved binaries less strongly than for the unresolved binaries. 
In the optimistic scenario for resolving binaries (Eq.\,\ref{eq:ang_sep_crit_pos}), the number of resolved WDMS and DWDs increases by 3--5\% compared to the standard scenario. In the pessimistic scenario, the number of resolved binaries decreases by about 10\%.
 Therefore, for resolved binaries the exact value of the critical angular separation is of little importance. Equally, the cut-off at $10^{10}$d for the lognormal distribution does not affect the number of resolved binaries significantly (about 1\%).

The observed number of resolved WDMS is in agreement with the lower limit of the models, and a factor of 2 below the upper limit (Table\,\ref{tbl:numbers_wds}, Fig.\,\ref{fig:nr}a). This is very similar to the case of single WDs. It indicates that our simulations and the adopted star formation histories are adequate in simulating space densities of the most common WD populations.

In contrast, the observed number of resolved DWDs is significantly lower than the predicted number, by a factor of 7--13. In other words, the BPS models predict 15--30 (isolated) resolved DWDs within 20\,pc, however, only two such systems are observed.

Regarding systems with high-order multiplicity, Table\,\ref{tbl:obs_ex} shows  two resolved WDMS in triples (WD0413$-$077 at 5\,pc and WD2351$-$335 at 22.9\,pc), and three triples with the WD as the outer companion (WD0433$+$270 at 18\,pc, WD0743$-$336 at 15.2\,pc, and WD1633$+$572 at 14.4\,pc).
Furthermore, there is a resolved DWD in a triple (WD0727+482) at 11.1\,pc (Table\,\ref{tbl:obs_ex}) and a DWD candidate with an MS companion at 17\,pc (WD2054$-$050). Including these systems does not significantly alter our conclusion.

\subsection{Discrepancy regarding resolved DWDs}
\label{sec:discr}
In this section, we investigate ways to resolve the discrepancy regarding the number of resolved DWDs between the simulations and observations, as found in the previous section.

\subsubsection{Non-isolated evolution}
\label{sec:dyn}
The binaries in our simulations are assumed to evolve in isolation, however, wide binaries can be significantly disturbed by dynamical interactions with, for example, other stars when passing through spiral arms, molecular clouds, or the Galactic tidal field \citep{Ret82, Wei87, Mal01, Jia10}. 
In extreme cases, these interactions can lead to the disruption of very weakly bound binaries.  
An observational limit to the semi-major axis in the Galactic disc is of the order of 0.1\,pc \citep[$5\cdot 10^6\,\Ro$, e.g.][]{Bah85, Clo90, Cha04, Kou07, Kou10}. 
Interesting to note in this context is our new DWD (Sect.\,\ref{sec:disc_0649}), which has a separation of $0.08\pm0.01$\,pc.  Systems with separations out to several parsec have been identified, although they are extremely rare \citep{Scholz08,caballero09,Mam10,Sha11}.
For models with the initial period distribution of \citet{Abt83}, there are no binaries with orbits wider than $5\cdot 10^6\,\Ro$, and roughly 15\% of resolved WDMS and 23\% of resolved DWDs are wider than $1\cdot 10^6\,\Ro$. For models with the lognormal distribution there are more wide binaries and the widest binaries are wider in comparison with the distribution of model `Abt'. 
The models with the lognormal distribution of periods predict that roughly 10\% (24\%) of resolved WDMS and 15\% (31\%) of resolved DWDs are wider then $5\cdot 10^6\,\Ro$ ($1\cdot 10^6\,\Ro$). 
If we assume that a binary will quickly dissolve once its orbit become larger than $5\cdot 10^6\,\Ro$ ($1\cdot 10^6\,\Ro$), the number of resolved binaries is reduced by $\lesssim$15\% ($\lesssim$30\%). 
This reduction is not sufficient to resolve the discrepancy between the observed and theoretical number of resolved DWDs. Also, the dissolution of a binary creates one or two single WDs, such that up to 14 (10--30) additional single WDs should be taken into account.  

\subsubsection{Stellar wind mass loss}
Another process that can lead to the disruption of a binary is a fast mass-loss event. In our simulations we have made the common assumption that the wind mass loss is slow compared to the orbital period. Within this limit, the change in the orbit becomes adiabatic, and the system remains bound \citep[see][for a review]{rahoma2009}. If, on the other hand, the mass loss is a sudden event, it  can  lead to the disruption of the system, as discussed in the context of supernova explosions \citep[e.g.][]{hills1983}. 
For a wide binary, a fast mass-loss phase can occur during the strong wind phases in the evolved stages of the star's evolution, such that the mass-loss interval is short compared to the orbital timescale \citep{hadjidemetriou1966, alcock1986, veras2011}.

As a proof of concept, we perform dynamical simulations of wind mass loss in binaries with four different mass ratios at a range of orbital separations (Appendix\,\ref{sec:app_wind}). 
We find that the majority of systems will not dissolve due to the stellar winds of their components. Only the orbits of the widest binaries ($a \gtrsim 10^6\,\Ro$) will indeed dissolve. The critical separation of order $10^6\,\Ro$ corresponds to systems in which the orbital period is comparable to the length of the asymptotic giant branch phase. As the critical separation for disruption by stellar winds is similar to that of dynamical interactions with Galactic objects, one can expect the effect on the population of wide, evolved binaries to be roughly similar to that discussed in Sect.\,\ref{sec:dyn}.

\subsubsection{Selection effects}
Another possible cause for our overestimation of wide binaries comes from the difficulty to identify binaries with large angular separations as bound objects. 
For the closest WDs in our sample ($\sim$3\,pc), the precise astrometry and the relatively few objects with similar distances mean that detection of nearby wide binaries is quite simple. 
However, for the most distant and faint objects, the relative errors on proper motions become much larger,
with the number of objects with consistent distances also increasing.
Therefore the detection of common proper-motion binaries at large distances becomes much more challenging.
To estimate the observational limit on the angular separation for discovering and confirming a proper motion pair, we inspect the WD proper-motion survey of \citet{farihi05}. In this sample, the angular separations for WD binaries range up to about 500" within 20\,pc. 
If we take 600" as the observational limitation, 
the number of binaries with the initial period distribution of model `Abt' would be reduced by $<5$\%, and for model `Lognormal' by $<20$\%. If instead we assume the sample is complete up to 100", the number of WD binaries decreases by 30--40\%. To conclude, this observational bias is not strong enough to explain the discrepancy in the observed and synthetic number of resolved DWDs.

\subsubsection{Binary formation}

To solve the discrepancy regarding DWDs, 
instead of a disruption, we consider the possibility that wide (zero-age MS) binaries are not formed as regularly as assumed in our models. 
We examine the effect on the WD space densities (of all types) of our modelling of the SFH, the initial period, and mass-ratio distribution.

The local SFH has been studied with a variety of techniques, and these studies have resulted in SFHs that range from constant values \citep[e.g.][]{Roc00, Rei07} to peaked distributions during the last $\sim$5Gyr \citep[e.g.][]{Ver02, Cig06, Tre14}.
If the majority of the star formation has taken place over the last few Gyrs, few low-mass stars would have had enough time to reach the WD stages of their evolution. 
As an experiment, we construct an alternative model similar to `cSFR' x `Abt', however, with a constant SFR only for the last 5\,Gyr, and no star formation at earlier times. As the absolute SFR in this model is arbitrary, we focus on the ratio of resolved WDMS to resolved DWDs.
Observationally there are 8.5 resolved WDMS for every resolved DWD, whereas the synthetic models predict ratios of 1.4--1.5 (which is another way of phrasing the discrepancy in the number of resolved DWDs between observations and models).   
In the experimental model, the ratio of resolved WDMS to DWDs increases to about 1.6. To conclude, a different model for the SFH that peaks at recent times can affect the total number of WDs, but does not resolve the discrepancy between theory and observations regarding the ratio of resolved WDMS and DWDs.

Regarding the distribution of initial periods, based on observations there are no indications that the distribution is dependent on the mass ratio of the system \citep[e.g.][]{Duc13}. Therefore, a different model of the initial period distribution is likely to affect the space density of resolved WDMS and DWDs equally, and therefore not solve the discrepancy in the number of resolved DWDs between observations and models.

Regarding the initial mass-ratio distribution, we examined the possibility that it is skewed towards unequal masses such that the companion star is of low mass and does not evolve far in a Hubble time. The observed mass-ratio distributions for different types of stars are approximately uniform down to $q\sim 0.1$ for $M\gtrsim 0.3\,\Mo$ \citep[see][for a review]{Duc13}. This is in support of our standard assumption of a uniform mass ratio distribution, however, we cannot exclude the possibility that the Galactic
stellar populations are not representative of the Solar neighbourhood.

\begin{figure}
    \centering
        \includegraphics[width=1.\columnwidth]{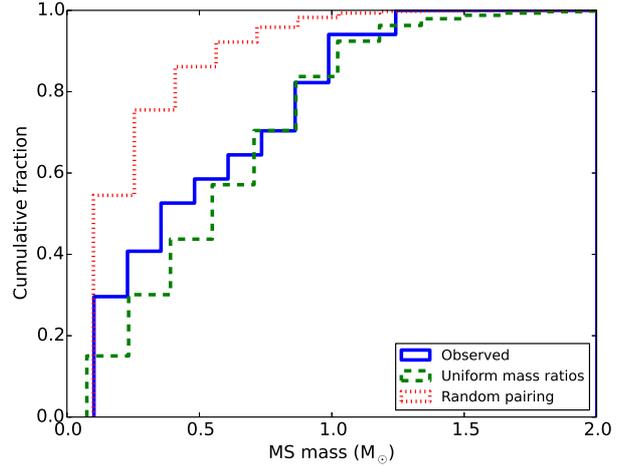}\\    
    \caption{Cumulative distribution of the mass of the MS components in resolved WDMS within 20\,pc. The observed spectral types from Table\,\ref{tbl:obs_bin} are converted to masses using \citet{Kra07}. The observed mass distribution is weighted according to the probability for each system to be within 20\,pc (blue solid line). 
    Both synthetic models shown are under the assumption of a constant SFR (i.e. model cSFR) and a log-uniform period distribution (i.e. model `Abt'), but differ in their treatment of the initial mass ratio distribution. The green dashed line represents the standard assumption in this paper of a uniform mass ratio distribution, whereas the red dotted line represents a random pairing of the primary and secondary mass. 
    }
    \label{fig:m_ms_wdms}
    \end{figure}

As an experiment, we constructed an alternative model to `cSFR' x `Abt', however, with an uncorrelated initial mass ratio distribution; that is, the masses of both stars are randomly drawn from the IMF. This significantly affects the number and ratio of resolved WDMS and resolved DWDs. Where our standard model predicts 22 unresolved WDMS and 15 unresolved DWDs, the experimental model predicts 60 unresolved WDMS and two unresolved DWDs. This mass-ratio distribution can be tested by comparing the synthetic and observed mass distribution of MS stars in resolved WDMS (Fig.\,\ref{fig:m_ms_wdms}). Our standard model (i.e. `cSFR' x `Abt') shows a uniform mass ratio distribution until about 1\,\Msolar, and a decline afterwards, as massive stars evolve into WDs. The observed mass distribution might indicate a slightly steeper distribution favouring low-mass companions, however, it is severely hampered by low-number statistics. With the current sample, a random-pairing of stellar masses in local binaries is excluded based on Fig.\,\ref{fig:m_ms_wdms}.

    \begin{figure}
    \centering
    \begin{tabular}{c}

        \includegraphics[width=1.\columnwidth, clip=true, trim =80mm 70mm 20mm 70mm]{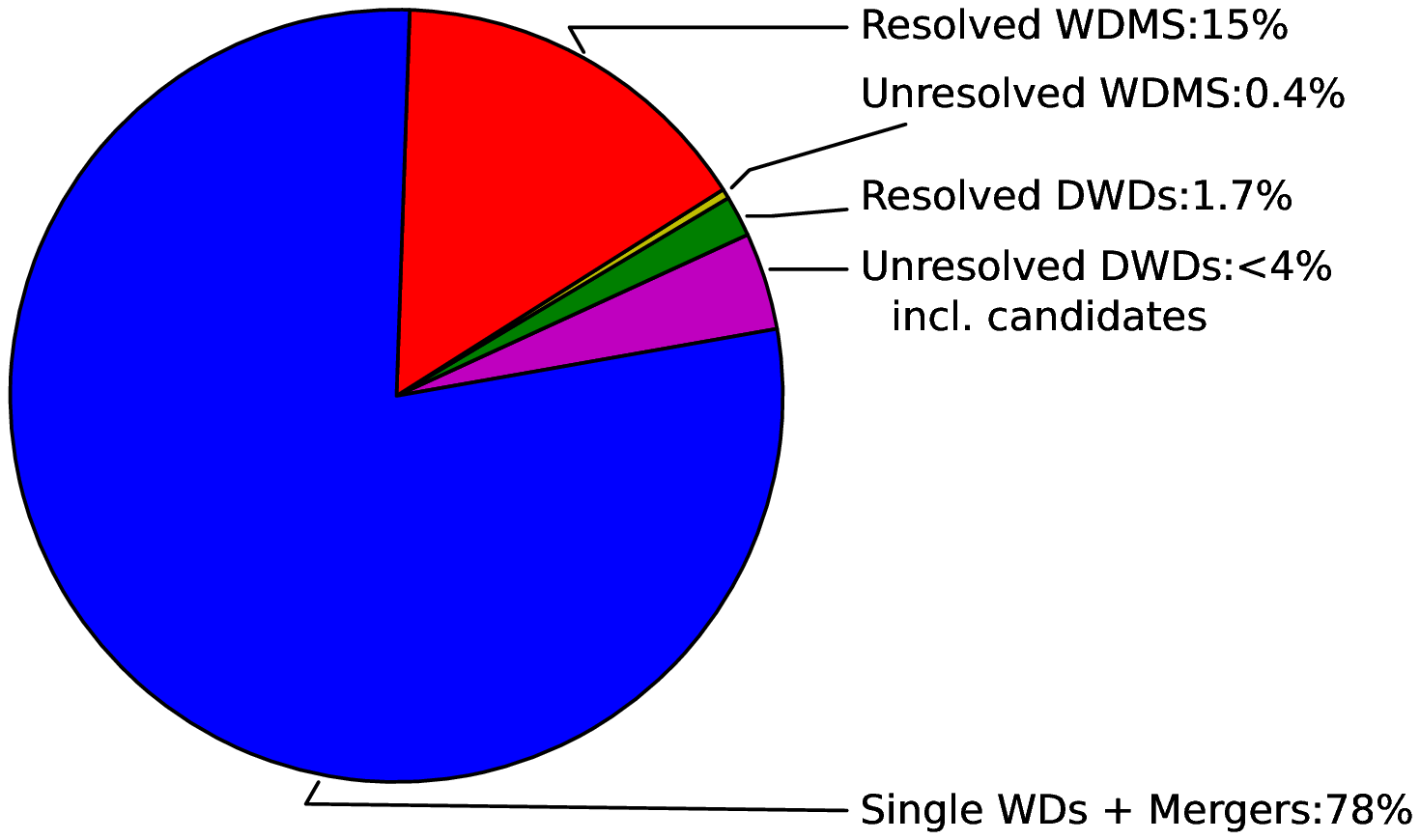} \\
        (a) Observed\\ \\
        \includegraphics[width=1.\columnwidth,clip=true, trim =80mm 70mm 20mm 70mm]{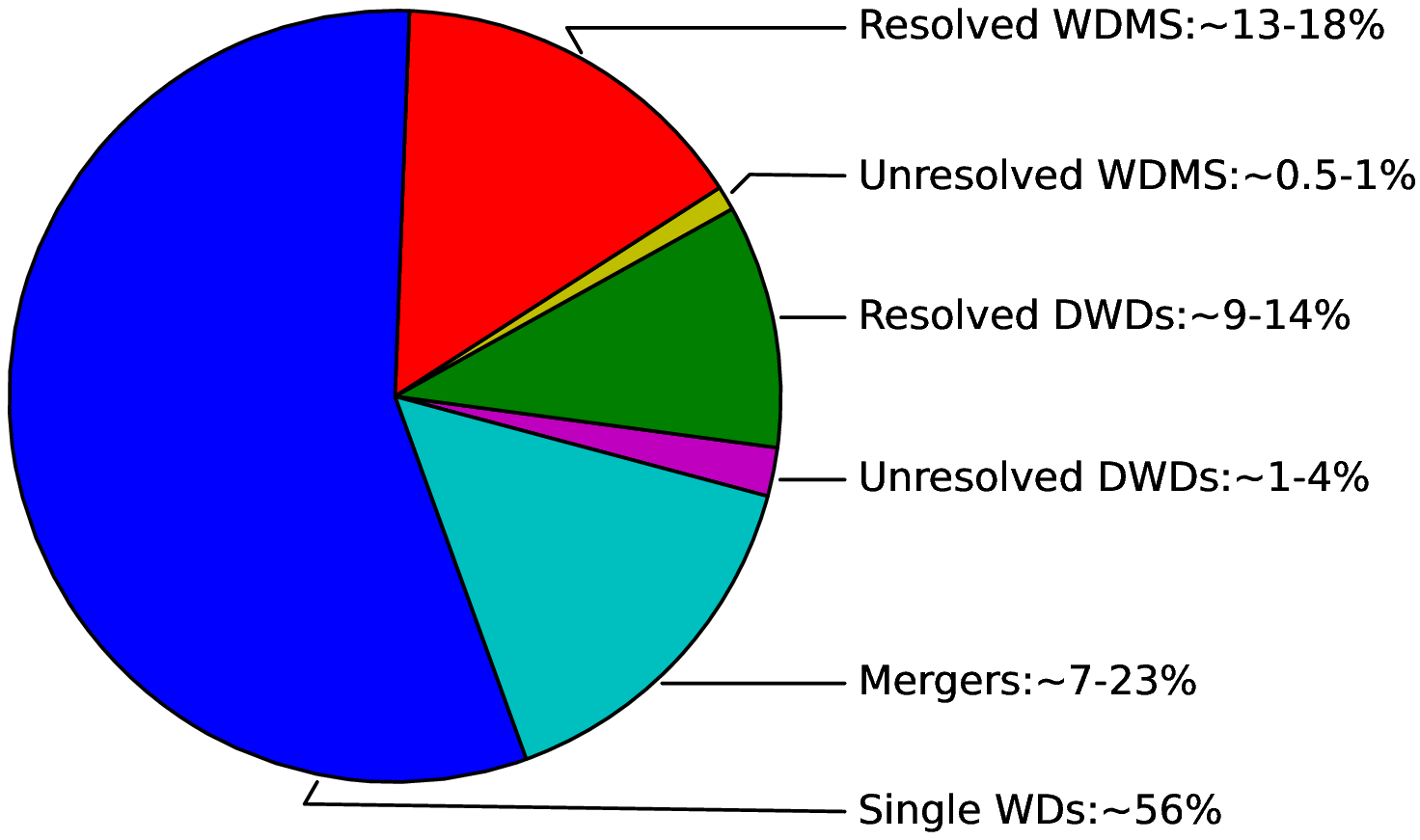} \\
         (b) Predicted\\
        \end{tabular}
    \caption{Current binary fraction for different WD systems. In the top panel the observed fractions are shown, on the bottom the range of fractions in the BPS models, based on Table\,\ref{tbl:numbers_wds}. From the BPS models the combination of single WDs and WDs from mergers should be compared with the observed single WDs. A significant discrepancy exists between observations and theory regarding resolved DWDs. }
    \label{fig:pie}
    \end{figure}

\subsection{The white dwarf binary fraction}
\label{sec:bin_fr_current}
The observed binary fraction amongst WDs in the 20\,pc sample ranges from 18--22\% depending on the binarity of the DWD candidates (Fig.\,\ref{fig:pie}a). If we include the  triple and quadruple systems, the observed fraction would be 22--26\%. This is in good agreement with the observed binary fraction of 26\% found by \citet{holberg09}. \citet{holberg09} focuses on the probability for a WD to be part of a binary or multiple star system, which is higher ($32\pm 8$\%) due to DWDs.

With our standard assumptions, we find a binary fraction for the 20\,pc WD population of about 25--35\% (Fig.\,\ref{fig:pie}b). 
For models with the period distribution of model `Abt', the binary fraction is slightly lower than for model `Lognormal', namely 23--28\% compared to 32--36\%. 
This is because in the lognormal distribution of initial periods, more wide binaries are formed which are less prone to merge during their evolution. 
If dynamical interactions or fast winds indeed disrupt wide binaries and create single WDs, the binary fraction decreases. For example for a semi-major axis limit of $10^6\,\Ro$, the binary fraction becomes 20--25\% and 26--30\% for model `Abt' and model `Lognormal', respectively. Therefore, if wide binaries are effectively destroyed, even the models with the lognormal initial period distribution give a binary fraction that is consistent 
with observations.

The \emph{current} binary fraction for the 20\,pc WD population is dependent on the initial (ZAMS) binary fraction, for which we have assumed a value of 50\%.\footnote{The difference between the initial and current binary fraction has been taken as evidence for missing binaries. See Sect.~\ref{sec:disc} for a discussion on this.} Observations have shown that the initial binary fraction is a function of the primary mass (Sect.\,\ref{sec:bin_fr}). 
Lowering the initial binary fraction to 40\% decreases the current binary fraction (see also Eq.\,\ref{eq:bf_ss}--\ref{eq:bf_bin}); 18--22\% and 25--28\% for model `Abt' and model `Lognormal', respectively. Similarly increasing the binary fraction to 60\%, increases the current binary fraction to 28--34\% and 39--45\% for model `Abt' and model `Lognormal', respectively. Unless wide binaries are very efficiently destroyed or the observations are very biased against finding common proper motion binaries, an initial binary fraction of 60\% gives a current binary fraction that is not in agreement with the observations. 
An initial binary fraction of 40--50\% is in agreement with observations of the average WD progenitor, that is, A-type stars \citep{derosa2014}.

\section{Outlook to Gaia}
\label{sec:gaia}

Gaia will have a strong impact on our understanding of Galactic stellar populations. 
The selection effects for the Gaia samples are clean and homogeneous, and therefore the samples will be very suitable for statistical investigations such as BPS studies.
Regarding WDs, Gaia is expected to increase the known sample significantly; from the current $\sim 2\cdot 10^4$ objects \citep{Kle13, Kep16} to a few $10^5$ WDs \citep{Tor05, Rob12, Car14}. 
In particular, the large sample size provides us with the opportunity to study rare WDs, for example WDs that are pulsating, magnetic, cool, part of the halo population, or possible supernova Type Ia progenitors. 
While the scientific potential of the WD sample has been discussed \citep[e.g.][]{Gae15}, little attention has been paid so far to WD binary systems.

In the Gaia era, the (relatively) complete sample of WDs is expected to extend from the current 20\,pc out to 50\,pc \citep{Car14, Gae15}.
Therefore, the effective volume of the complete WD sample increases by more than an order of magnitude.
We predict the number of single and binary WDs within 50\,pc (see Table\,\ref{tbl:numbers_wds_gaia}) with a BPS approach similar to that used previously in Sect.\,\ref{sec:res} and described in Sect.\,\ref{sec:bps}. Our model of the selection effects for the 50\,pc sample are specific to the Gaia sample, and described below.

\subsection{Single white dwarfs with Gaia}
Within 50\,pc, Table\,\ref{tbl:numbers_wds_gaia} shows that we expect to detect thousands of single WDs.
This vast number of single WDs in a volume-limited sample will allow for an accurate determination of the luminosity function and the mass function, which will not be affected by brightness-related selection effects. 
These studies have the potential to teach us about the SFH, initial-final mass relation for WDs, and the initial mass functions of WDs. Furthermore, our models show that several hundreds of single WDs will be detected that formed through a merger in a binary system. With an increasingly more detailed analysis of the complete WD sample, it will become important to understand how to distinguish merged objects from single stars that evolved completely isolated.

\subsection{Resolved binaries with Gaia}
Due to its high precision astrometry, the Gaia mission is very proficient in the detection of binaries and systems of higher-order multiplicities.
The high precision astrometry leads to improved proper motions, and parallax measurements with uncertainties of $\sim$1\% for WDs within 100\,pc \citep{Car14}. 
This is particularly important for the detection of resolved binaries, which can be identified either by their common proper motion (and distance) or astrometrically. 

The capability of Gaia to resolve a system into two local maxima depends on the angular separation, magnitude difference between the two stars, and the orientation angle of the binary orbit with respect to the scan axis of Gaia \citep{DeB15}.    
We assume the critical separation for resolving two stars is:
\begin{equation}
\mathrm{log}(s_{\rm crit, gaia}) = 0.075|\Delta G|-0.53,
\label{eq:ang_sep_crit_gaia}
\end{equation}
where $|\Delta G|$ is the difference in the $G$-band magnitude of the two stellar components.
The functional form of Eq.\,\ref{eq:ang_sep_crit_gaia} is very similar to the one we derived for the 20\,pc sample in Eq.\,\ref{eq:ang_sep_crit}. 
 The Gaia $G$-band magnitudes are calculated using the formalism of \citet{jordi10}.
Equation\,\ref{eq:ang_sep_crit_gaia} is a fit to the results of \citet{DeB15}, who calculate the probability for resolving two stars with Gaia as a function of angular separation and magnitude difference, averaged over all orientation angles (their Figs.\,18~and~19). 
Gaia’s resolving power does not vary with the magnitude of the primary for a given $|\Delta G|$.
Equation\,\ref{eq:ang_sep_crit_gaia} is a fit to the contour of 50\% probability, with the idea that the system will be resolved in at least one of the transits observed during the mission. 

Overall, the critical angular separation is about 0.3\,arcsec. This is a vast improvement compared to that of the current sample (Sect.\,\ref{sec:types}), and as a result one would expect the ratio of resolved binaries to unresolved binaries to increase compared to that of the 20\,pc sample. However, as the typical distances for the Gaia sample are larger than for the 20\,pc sample, the relative number of resolved binaries remains approximately the same. The total number of resolved binaries is very similar to what is expected from solely the increase in effective volume, which gives an increase of about a factor of 15.

In absolute numbers, Table\,\ref{tbl:numbers_wds_gaia} shows that the BPS models predict that hundreds of resolved binaries can be observed within 50\,pc. The Gaia sample is, therefore, expected to overcome the small number statistics by which the 20\,pc sample is hampered. 
Consequently, the Gaia sample will shed more light on the current discrepancy between the observations and models regarding the space density of resolved DWDs. Additionally, the sample of resolved WDMS will expose the initial mass-ratio and period distribution of wide binaries, and show if these can resolve the just mentioned discrepancy. Lastly, the widest binaries with separation above $10^5$\,\Rsolar\, will give insights into the formation of wide binaries.

\subsection{Unresolved binaries with Gaia}

Unresolved binaries can be recognized within the Gaia data based on their odd colours, odd absolute magnitudes, or due to their poor fit to an astrometric model of a single star. Regarding the colours, we model the selection effects in a way similar to Eqs.\,\ref{eq:g_bias}-\ref{eq:r_bias}, but based on Gaia colours. This has the advantage that it guarantees that the relevant photometry and astrometry is available for all stars in
a homogeneous way. We assume that unresolved WDMS can be recognized as a binary when:  
\begin{equation}
\Delta G_{\rm BP} \equiv G_{\rm BP, WD}-G_{\rm BP, MS}<1,
\label{eq:Gbp_bias}
\end{equation}
and 
\begin{equation}
\Delta G_{\rm RP} \equiv G_{\rm RP, WD}-G_{\rm RP,  MS} > -1, 
\label{eq:Grp_bias}
\end{equation}
where $G_{\rm BP}$ and $G_{\rm RP}$ represent the magnitudes in the Gaia BP and RP bands for the WD and MS component. For DWDs, we require: 
\begin{equation}
\Delta G \equiv |G_{\rm WD1}-G_{\rm WD2}|<1. 
\label{eq:G_bias}
\end{equation}

Alternatively binaries could be detected by their odd absolute magnitude.
If the photometric or spectroscopic distance is significantly different from the 
trigonometric distance, the system can be flagged as a binary candidate. 
Due to the high precision astrometry of Gaia, the error on the trigonometric distance is negligible. Assuming a 10\% accuracy for the WD spectroscopic distances, there would be a discrepancy with the trigonometric distance if:
\begin{equation}
G_{\rm WD, bright} - G_{\rm total} > -5\, {\rm log\,} 0.9,
\label{eq:dist_bias_DWD}
\end{equation}
where $G_{\rm WD, bright}$ and $G_{\rm total}$ are the $G$-band magnitude of the brightest WD component and that of the binary as a whole, respectively. This is equivalent to $\Delta G <1.57 $ (see also Eq.\,\ref{eq:G_bias}). As the mass-radius relationship is less strict for MSs than for WDs, we assume the accuracy for the distance determination to MSs is lower, that is, 20\%.
An MS can be discovered to host a companion, if: 
\begin{equation}
G_{\rm MS} - G_{\rm total} > -5\, {\rm log\, }0.8,
\label{eq:dist_bias_WDMS}
\end{equation}
where $G_{\rm MS}$ is the $G$-band magnitude of the MS.

Similar to single WDs and resolved WD binaries, the largest volume-limited sample of unresolved WD binaries is about 15 times as large as the current sample (Table\,\ref{tbl:numbers_wds_gaia}). The BPS models predict that about 10--30 unresolved WDMS and 20--130 unresolved DWDs can be observed with Gaia. For the visible WDMS, 94\% of the systems are selected based on their odd colours; that is, these systems fulfil Eqs.\,\ref{eq:Gbp_bias}~and~\ref{eq:Grp_bias}. Similarly for the visible DWDs, the majority of binaries have odd colours; 90\% for models $\mga$ and $\maa$, and 84\% for model $\maa$2. 
Assuming that accurate periods can be determined by the radial velocity method up to ten days, the number of close DWDs with known periods are reduced to less than ten for model $\maa$2, a few tens for model $\maa$, and several tens for model $\mga$ (last column Table\,\ref{tbl:numbers_wds_gaia}).
These DWDs will be extremely useful to constrain the CE-phase, for example by modelling the specific evolution of each system as in \citet{Nel00}. Furthermore, as the number of unresolved DWDs (with and without known periods) in the complete 50\,pc sample is strongly dependent on the modelling of the CE-phase, the number of systems provides an extra constraint for the CE-phase.

Lastly, unresolved astrometric binaries can be recognized from their poor fit to a standard single star astrometric model. For many it should be possible to determine a photocentre orbit with semi-major axis $a_{\rm photo}$ \citep{gontcharov02, sahlmann15}:
\begin{equation}
a_{\rm photo} = \Big( \frac{M_{\rm faint}}{M_{\rm bright} + M_{\rm faint}} -  \frac{L_{\rm faint}}{L_{\rm bright} + L_{\rm faint}} \Big)\, a, 
\label{eq:a_photo}
\end{equation}
where $L_{\rm bright}$ and $L_{\rm faint}$ are the luminosities of the bright and faint stellar component. A common detection criterion for astrometric binaries is $a_{\rm photo} / \sigma > 3$ \citep[e.g.][]{casertano08, sozzetti14}, where $\sigma$ is the astrometric precision of Gaia \citep{DeB14}. The precision is a function of the $G$-band magnitude and the $V-I$-colour of the system, where for the latter we use the transformations of \citet{jordi06}. For faint sources, such as the WDs in our sample, the precision is a few hundred $\mu$as. From the astrometric motion of the binary photocentre, it will be possible to derive the orbital period, however, it will be difficult to work out the nature of the unseen companion of the unresolved binary. 
For WD primaries with an astrometric perturbation, there is a good possibility that the companion is a WD as well, and therefore we focus on DWDs. 
The BPS models predict 20--45 unresolved astrometric DWD binaries within 50\,pc. The majority of these have orbital separations just below $s_{\rm crit, gaia}$. Only three--six DWDs are compact enough to have experienced one or more phases of mass transfer during their formation. If an unresolved astrometric DWD is observed for which both masses can be measured spectroscopically, it would be a very interesting system to constrain CE-evolution, in particular because the astrometric method to determine periods is sensitive to longer periods than is feasible with the spectroscopic method.

\begin{table*}
\caption{Number of systems with WDs components within 50\,pc for different BPS models. The Gaia WD sample is expected to be roughly complete out to approximately 50\,pc (Sect.\,\ref{sec:gaia}). Here a limiting angular separation of 0.3" is assumed to differentiate between resolved and unresolved binaries. The table layout is the same as Table\,\ref{tbl:numbers_wds} with one extra column. The column on the far right shows the number of unresolved DWDs with periods less than ten days. The statistical error is omitted, as it is smaller than the systematic error, that is, variation between the different BPS models.}
\centering
\begin{tabular}{lll|ccccccc}
\hline
\hline
\multicolumn{10}{c}{\textbf{BPS models} }  \\
\hline
 && & \multirow{2}{*}{Single stars} & \multirow{2}{*}{Mergers} & \multicolumn{2}{c}{WDMS}  & \multicolumn{3}{c}{DWD}  \\
SFH & Period distr. & CE &   &  & Resolved  & Unresolved & Resolved  & Unresolved & P$<$10d\\
\hline
\hline
\multirow{3}{*}{BP} & \multirow{3}{*}{Abt} & $\mga$ & \multirow{3}{*}{1884} &556&\multirow{3}{*}{445}&28&\multirow{3}{*}{316}&132 & 86\\
 &  & $\maa$ & &640 && 31&&65 & 38\\
 &  & $\maa2$ & &777&&16&&33 & 7.7\\
\hline
\multirow{3}{*}{BP} & \multirow{3}{*}{Lognormal} & $\mga$ &\multirow{3}{*}{1884} &239&\multirow{3}{*}{599}&30&\multirow{3}{*}{440}&126 & 73\\
 &  & $\maa$ & &297&&32&&65 & 43 \\
 &  & $\maa2$ & &427&&19&& 38& 8.1\\
\hline \hline
\multirow{3}{*}{cSFR} & \multirow{3}{*}{Abt} & $\mga$ & \multirow{3}{*}{1389}&406&\multirow{3}{*}{330}&22&\multirow{3}{*}{234}& 100 & 68\\
 &  & $\maa$ & &467&&23&&50 & 31\\
 &  & $\maa2$ & &588&&12&& 22 & 5.4\\
\hline
\multirow{3}{*}{cSFR} & \multirow{3}{*}{Lognormal} & $\mga$ &\multirow{3}{*}{1389} &177&\multirow{3}{*}{446}&24&\multirow{3}{*}{327}& 93 & 56\\
 &  & $\maa$ & &219&&24&&48 & 26 \\
 &  & $\maa2$ & &328&&15&& 26 & 5.6\\
\hline
\end{tabular}
\\
\label{tbl:numbers_wds_gaia}
\end{table*}

\section{Discussion on missing binaries}
\label{sec:disc}

\citet{ferrario12} noted a tension between the high binary fraction of Solar-type MS stars (here initial binary fraction, $\sim$50\%) and the low binary fraction of WDs (here current binary fraction, $\sim$25\%). Due to this discrepancy, they advocate there must be an additional $\sim$25\% of as yet undiscovered WDs hiding in unresolved binaries. 
However, we find that when taking into account the full binary evolution and including selection effects, this tension is largely removed. The dominant reason in most BPS models is that the binaries may merge during their evolution. A secondary reason is that a WD may hide in the glare of the primary star. In our models, for every (detectable) unresolved WDMS, there are eight WDMS systems that would not be recognized as a WDMS due to the luminosity contrast.

Another claim of missing binaries with WD components has come from \citet{katz14}, based on the luminosity function of the resolved WDMS in the 20\,pc sample. With a similar reasoning as \citet{ferrario12}, \citet{katz14} argue there is a deficit of up to 100 WDs in binary systems within 20\,pc. They conclude that it is likely that the number of WDMS is roughly equal to or higher than that of single WDs. This conclusion is not supported by our results; we find approximately five times as many single WDs (both from single stellar evolution as from binary mergers) as WDMS, which is consistent with the observations (Table.\,\ref{tbl:numbers_wds}).

Beyond 20\,pc, it has been claimed by \citet{holberg09} that a significant number of Sirius-like systems (resolved WDMS with companions of spectral type K or earlier) are missing. This is based on a comparison of space densities at different distances from the Sun. A comparison with BPS models is outside the scope of this paper.

\subsection{Resolved DWDs}
\label{sec:disc_dwd}
Our simulations show a discrepancy with the observations for the number of resolved DWDs. 
The BPS models predict a factor of 7--13 more systems than what is observed. This large factor is remarkable as resolved binaries are too wide for mass transfer to take place. The stars have practically evolved as if they were isolated stars. Therefore, there are only a few physical processes that affect the number density of resolved binaries.

The (apparent) disruption of wide binaries is not likely to solve the discrepancy. We considered disruptions due to dynamical interactions with other stars, molecular clouds, or the Galactic tidal field, and due to stellar winds that are short-lived compared to the binary period. In addition, we studied the apparent disruption of wide binaries from selection biases against finding common proper motion pairs. 

It is possible that the progenitors of wide DWDs are not as commonly formed as previously assumed. We considered three options: 
\begin{itemize}
\item The star formation rate and initial stellar space density are likely not the cause for the discrepancy, as the space density of single WDs and resolved WDMS are modelled correctly within a factor of 2. 
\item The binary fraction decreases as the primary mass increases. In this case, fewer binaries with massive stars are born that can form WDs in a Hubble time. This does not seem likely as the binary fraction is observed to increase with primary mass \citep[e.g.][]{Duc13}, however, we cannot discard the possibility that locally it could be different.
\item In this study we have assumed a uniform mass-ratio distribution for the ZAMS-binaries, which is the current consensus among surveys of different types of field stars \citep[e.g.][]{Duc13}. 
However, there are observational \citep{Rag10, derosa2014} indications that the mass-ratio distribution of close and wide binaries are distinct and that for wide binaries ($>125$ AU) the distribution tends towards unequal masses.
In this scenario, the companion stars are biased to low masses and would not evolve far in a Hubble time. This would decrease the number of expected DWDs, but increase the number of WDMS. 
Even though the 20\,pc sample is severely hampered by small number statistics, the mass distribution of the MS-component of resolved WDMS might indicate a mass-ratio distribution that is slightly steeper than uniform, that is, one which favours low mass companions. Our BPS models predict that the small number statistics can be overcome with the 50\,pc sample based on Gaia (Table\,\ref{tbl:numbers_wds_gaia}).
\end{itemize}

The last option we consider is that at least ten resolved DWD systems have been missed observationally.  The chance that this is due to Poisson fluctuations is less than 0.005\%.

\section{Conclusion} 
\label{sec:concl}
The sample of white dwarfs within 20\,pc of the Sun is extraordinary due to its high level of completeness of 80--90\%. It is also relatively unbiased with respect to WD luminosity and cooling. 
From a literature study, we compiled the most up-to-date sample and divided it into different binary types. We compared the sample with the results of a binary population synthesis study in which the evolution of binaries is modelled starting from the zero-age main-sequence. Where many  BPS studies focus on a single binary population, the 20\,pc sample allows for a consistent and simultaneous study of the six most common WD systems. 
Moreover, the 20\,pc sample allows for a strong test on the synthetic space density estimates of the local WD populations, and in turn the synthetic event rates and space density estimates of other stellar populations as well.

We have constructed (2x2x3=) 12 BPS models that differ in their treatment of the SFH, initial period distribution of the binaries, and the CE-phase for interacting binaries. 
The statistical error on the BPS results is small, for example the uncertainty on the space densities is $<10\%$ . The main source of uncertainty in BPS simulations comes from the uncertainty in the input assumptions \citep[and not from numerical effects, see also][]{Too14}:  
\begin{itemize}[noitemsep,topsep=0pt]
\item The different models of the SFH affect the WD space densities by $\sim$50\%. 
\item The different models of the initial binary period distribution affect most strongly the space densities of single WDs that are formed through mergers of binary systems. It affects their space density by a factor of $\sim$2. 
\item The space densities of unresolved binaries are most strongly affected by the uncertainty in the common-envelope phase,  by about a factor of 2 and 4, for WDMS and DWDs respectively. \\
\end{itemize}

Our main results can be summarized as follows:
\begin{itemize}[noitemsep,topsep=0pt]
\item Overall, we find that the number of systems predicted by the BPS models for the different types of WD systems are in good agreement with the observations. 
We show that the BPS estimates of the number of WDs within 20\,pc are well calibrated, which gives confidence in the synthetic space densities and event rates for other populations.

\item With an initial binary fraction of 50\%, the number of observed and predicted single WDs and resolved WDMS agrees within a factor of 2. 
This may indicate that the local star formation rate is somewhat overestimated, in particular model BP where the model of the Galaxy is based on \citet{Boi99}. In this model of the Galactic history, star formation has proceeded for 13.5\,Gyr in the disc, however from MS and WD populations in the Galactic disc a maximum age of 8--10\,Gyr seems more appropriate. 

\item 
We find that the number of single WDs that are formed from mergers in binaries is significant, about 10--30\%. Therefore, it is important to take mergers into account in studies that derive the SFR and initial mass function from observed WD samples.

\item
Regarding the space densities of unresolved binaries, we find that the BPS models are consistent with the observations, however, the errors on both measurements are large. 
The main source of uncertainty on the synthetic numbers comes from the uncertainty in the common-envelope phase and the modelling of the selection effects. The observations are hampered by low number statistics and the fact that the binarity is not confirmed for all DWD candidates. 
Larger number statistics, such as expected for Gaia,  would allow for stronger constraints on the BPS models.

\item We find a discrepancy between the observed and synthetic number of resolved DWDs. Our models overpredict the number of resolved DWDs by a factor of 7--13. We have studied several possible mechanisms for the (apparent) disruption of wide binaries, but show that these are not likely to solve the discrepancy (Sect.\,\ref{sec:discr}). 
Either more than ten resolved DWDs have been missed observationally in the Solar neighbourhood, or the initial mass-ratio distribution is biased towards low-mass ratios, of which there are some indications in the 20\,pc sample (see also Sect.\,\ref{sec:disc_dwd} for a full discussion).

\item 
We predict the number of single and WD binary systems within 50 \,pc of the Sun. This is the largest volume-limited sample that can be fully observed by Gaia. 
 We predict it will contain thousands of single WDs, hundreds of single WDs that are formed due to a merger in a binary, hundreds of wide binaries, and several dozen  unresolved binaries.  
The large data set of single WDs allows for detailed studies of e.g. the space density, mass function, and luminosity function.  The large population of wide binaries in the 50\,pc sample can provide stringent tests of WD evolutionary models, for example the age of the stellar components, the initial-final mass relation of WDs, or the mass-radius relation of WDs, and in particular the discrepancy between the observed and synthetic number of resolved DWDs. 
The population of resolved and unresolved binaries can provide additional information, for example on the  period- and mass-ratio distributions of the WD binaries. As such the 50\,pc sample has the potential of breaking the degeneracy between the synthetic models. 
\end{itemize}

\begin{acknowledgements}
We thank Elme Breedt, Roberto Raddi, and Silvia Catalan for the discussions on WDs, Pierre Maxted \& Tom Marsh for the discussions on wide binaries, and Carmen Martinez Barbosa, Gr\'ainne Costigan, and Anthony Brown for the discussions on Gaia. This work was supported by the Netherlands Research Council NWO (grant VIDI [\# 639.042.813]), by the Netherlands Research School for Astronomy (NOVA), and the Frye stipend of the Radboud University Nijmegen. 
The research leading to these results has also received funding from the
European Research Council under the European Union's Seventh Framework
Programme (FP/2007-2013) / ERC Grant Agreement n. 320964 (WDTracer).

\end{acknowledgements}

\begin{appendix}

\section{Sample of observed single WDs}
Table\,\ref{tbl:obs_single} shows the sample of observed single WDs. This is mainly based on \citet{Gia12}, with additions of \citet{limoges13},  \citet{sion14}, and \citet{limoges15}. 

\label{sec:app_sample}

\begin{table*}
\caption{Known single WDs in the Solar neighbourhood. This sample is mostly based on \cite{Gia12} with additions and modifications from papers indicated in the last column. }
\centering
\begin{tabular}{lccccc}
\hline
\hline
 & Distance [pc] & Spectral type & Mass [\Msolar]  & log L/\Lsolar & References \\
\hline
0000$-$345     & 13.2 (1.6) & DAH & 0.88 (0.10) & -3.82 &       \\
0004+122     & 21.0 (3.4)&  & 0.57 (0.15) & -4.02 &                             1\\
0005+395 &      20.21   (1.25)  &       &       0.58    (-)&    -& 2\\
0008+424     & 21.4 (1.1) & DA& 0.64 (0.04) & -3.45 &         \\    
0009+501     & 11.0 (0.5) & DAP& 0.73 (0.04) & -3.72 &         \\
0011$-$134     & 19.5 (1.5) & DAH& 0.72 (0.07) & -3.85 &       \\
0011$-$721     & 17.6 (0.7) & DA& 0.59 (0.00) & -3.63 &       \\
0019+423     & Sect.\,\ref{sec:distance}&   & 0.58 (0.15) & -3.85 & 1\\
0025+054 &      21.12   (1.71) &                &       0.58    (-)&    -& 2\\
0038$-$226     & 9.05 (0.10) & DQpec& 0.53 (0.01) & -3.94 &      \\
0046+051     & 4.297 (0.033) & DZ&  0.68 (0.02) & -3.77 &      3, 4, 5\\
0053$-$117     & 20.7 (1.3) & DA& 0.67 (0.05) & -3.49 &       \\
0115+159     & 15.4 (0.7) &DQ & 0.69 (0.04) & -3.1 &          \\
0123$-$262     & 21.7 (0.8) &DC & 0.58 (0.00) & -3.4 &        \\
0136+152     & 21.2 (0.8)&  & 0.72 (0.03) & -3.34 &                             1\\
0141$-$675     & 9.73 (0.080) & DA& 0.48 (0.06) & -3.55 &     6\\
0148+467     & 15.5 (0.8) & DA& 0.63 (0.03) & -2.26 &   2, 7, 8 \\
0208+396     & 16.7 (1.0) & DAZ& 0.59 (0.05) & -3.39 &         \\
0213+396     & 20.9 (0.9) & DA& 0.8 (0.03) & -3.14 &          \\
0213+427     & 19.9 (1.6) & DA & 0.64 (0.08) & -3.93 &         \\
0230$-$144     & 15.6 (1.0) & DA& 0.66 (0.06) & -3.96 &       \\
0233$-$242     & 16.7 (0.7) & DC& 0.58 (0.00) & -3.94 &       \\
0236+259     & 21.8 (0.8) & DA& 0.59 (0.00) & -3.83 &         \\
0243$-$026     & 21.2 (2.3) & DAZ& 0.7 (0.10) & -3.62 &        \\
0245+541     & 10.3 (0.3) & DAZ& 0.73 (0.03) & -4.13 &         \\
0252+497 &      17.99   (2.9)    &  &   1.2      (0.11)&                -& 2\\
0255$-$705     & 27.8 (1.1) & DA& 0.57 (0.03) & -2.67 &       \\
0310$-$688     & 10.15 (0.15) &DA & 0.67 (0.03) & -1.97 &     2, 3, 4, 5 \\
0322$-$019     & 16.8 (0.9) & DAZ& 0.63 (0.05) & -4.02 &       \\
0340+198 &      19.5    (0.83) &                &               0.94    (0.05)&         -& 2\\
0341+182     & 19.0 (1.1) & DQ& 0.57 (0.06) & -3.57 &         \\
0344+014     & 20.6 (1.2) & DC& 0.58 (0.00) & -3.99 &         \\
0357+081     & 17.8 (1.2) & DA& 0.61 (0.06) & -3.91 &         \\
0414+420 &      23.8    (3.6)    &      &               0.58    (-)&    -& 2\\
0423+044     & 20.9 (1.7)&  & 0.67 (0.08) & -4.22 &             9, 10 \\
0435$-$088     & 9.51 (0.24)& DQ & 0.53 (0.02) & -3.59 &      \\
0457$-$004     & 28.7 (1.4) & DA& 1.07 (0.03) & -3.09 &       \\
0511+079     & 20.3 (0.6)&  & 0.8 (0.08) & -3.75 &                      9, 10, 11\\
0541+620 &      20.4    (3.2 ) &                &               0.58    (-)&    -& 2\\
0548$-$001     & 11.1 (0.3) & DQP& 0.69 (0.03) & -3.8 &        \\
0552$-$041     & 6.412 (0.032) & DZ& 0.82 (0.01) & -4.21 &    3, 5, 6\\
0553+053     & 8.0 (0.23) & DAH & 0.72 (0.03) & -3.91 &        \\
0618+067     & 22.6 (2.1)&  & 0.93 (0.17) & -4.05 &                 1\\
0620$-$402     & 25.3 (4.0)&  & - & - &                                         9, 12\\
0644+025     & 18.4 (1.9)&  DA& 1.01 (0.07) & -3.79 &         \\
0644+375     & 15.276 (0.423)&  DA& 0.69 (0.03) & -1.48 &     3, 4, 5\\
0655$-$390     & 17.1 (0.7)&  DA& 0.59 (0.00) & -3.64 &       \\
0657+320     & 18.7 (0.3)& DA & 0.6 (0.02) & -4.1 &           \\
0659$-$063     & 12.3 (1.3)& DA & 0.82 (0.07) & -3.77 &       \\
0708$-$670     & 17.3 (0.6)& DC & 0.57 (0.00) & -4.02 &       \\
0728+642     & 18.4 (0.5)&  DAP& 0.58 (0.00) & -4.0 &          \\
0749+426     & 24.6 (0.8)& DC & 0.58 (0.00) & -4.2 &          \\
0752$-$676     & 7.898 (0.082)& DA & 0.73 (0.06) & -3.94 &    3, 5, 6\\
0802+387     & 20.8 (1.8)&  & 0.73 (0.02) & -4.13 &                 1\\
0805+356     & 24.5 (0.8)&  & 0.83 (0.03) & -3.25 &             9, 13\\
0806$-$661     & 19.2 (0.6)& DQ & 0.58 (0.03) & -2.80 &       \\
0810+489     & 18.3 (0.6)& DC & 0.57 (0.00) & -3.55 &         \\
0816$-$310     & 22.1 (1.6)& DZ & 0.57 (0.00) & -3.61 &       \\
0821$-$669     & 10.7 (0.1)& DA & 0.66 (0.01) & -4.08 &       \\
0827+328     & 22.3 (1.9)& DA & 0.84 (0.07) & -3.64 &         \\
0840$-$136     & 13.9 (0.8)& DZ & 0.57 (0.0) & -4.1 &         \\
0843+358     & 27.0 (1.5)& DZA & 0.58 (0.0) & -3.02 &          \\
0856+331     & 20.5 (1.4)& DQ & 1.05 (0.05) & -3.32 &         \\
0912+536     & 10.3 (0.2)& DCP & 0.75 (0.02) & -3.57 &         \\
0939+071     & 18.9 &  & - & - &                            \\
0946+534     & 23.0 (1.9)& DQ & 0.74 (0.08) & -3.35 &         \\
0955+247     & 24.4 (2.7)& DA & 0.76 (0.10) & -3.24 &         \\
\hline
\end{tabular} 
\label{tbl:obs_single}
\end{table*}

\begin{table*}
\renewcommand\thetable{2b}
\centering
\begin{tabular}{lccccc}
\hline
\hline
 & Distance [pc] & Spectral type & Mass [\Msolar]  & log L/\Lsolar  & References \\
\hline
1008+290     & 14.8 (0.1)& DQpecP & 0.68 (0.01) & -4.31 &         \\
1019+637     & 16.4 (1.0)& DA & 0.57 (0.05) & -3.5 &          \\
1033+714     & 19.6 (0.8)& DC & 0.58 (0.00) & -4.15 &         \\
1036$-$204     & 14.3 (0.1)& DQpecP & 0.6 (0.01) & -4.19 &        \\
1055$-$072     & 12.2 (0.5)& DC & 0.85 (0.04) & -3.6 &        \\
1116$-$470     & 17.5 (0.7)& DC & 0.57 (0.00) & -3.8 &        \\
1121+216     & 13.4 (0.5)& DA & 0.71 (0.03) & -3.46 &         \\
1124+595     & 27.6 (1.3)& DA & 0.98 (0.03) & -3.09 &         \\
1134+300     & 15.3 (0.7)& DA & 0.97 (0.03) & -1.78 &         \\
1142$-$645     & 4.634 (0.008) & DQ & 0.61 (0.01) & -3.27 &    3, 4, 5, 6, 7, 8\\
1143+633     & 21.3 (3.4)&  & 0.58 (0.15) & -3.95 &                 1, 14\\
1145-451 &      22.94   (2.08) &                &       0.58    (0.12)&          -& 2\\
1148+687     & 18.0 (0.6)&  & 0.69 (0.04) & -3.64 &             9, 15\\
1202$-$232     & 10.83 (0.11)& DAZ & 0.59 (0.03) & -3.05 &     6\\
1208+576     & 20.4 (1.9)& DAZ & 0.56 (0.09) & -3.74 &         \\
1223$-$659     & 10.26 (0.31)& DA & 0.45 (0.02) & -3.16 &     6\\
1236$-$495     & 16.4 (2.6)& DAV & 1.0 (0.11) & -2.97 &        \\
1257+037     & 16.6 (1.0)& DA & 0.7 (0.06) & -3.95 &          \\
1309+853     & 16.5 (0.3)& DAP & 0.71 (0.02) & -4.01 &         \\
1310+583     & 24.9 (1.0)& DA & 0.66 (0.03) & -2.77 &         \\
1310$-$472   & 15.0 (0.5)&  DC & 0.63 (0.04) & -4.42 &         \\
1315$-$781   & 19.2 (0.3)& DC & 0.69 (0.02) & -3.94 &         \\
1334+039     & 8.24 (0.23)& DA & 0.54 (0.03) & -4.02 &        \\
1339$-$340   & 21.0 (1.2)& DA & 0.58 (0.00) & -3.96 &         \\
1344+106     & 20.0 (1.5)& DAZ & 0.65 (0.07) & -3.49 &         \\
1344+572     & 25.8 (0.8)&  & 0.53 (0.03) & -4.02 &             9, 11\\
1350$-$090   & 25.3 (1.0)& DAP & 0.68 (0.03) & -2.98 &         \\
1425$-$811   & 26.9 (1.0)&  DAV& 0.61 (0.03) & -2.46 &         \\
1443+256 &      17.5    (2) &           &               0.58    (-)&    -& 2\\
1444$-$174   & 14.5 (0.8)& DC & 0.82 (0.05) & -4.27 &         \\
1524+297 &      22.4    (2.6) &         &               0.58    (-)&    -& 2\\
1532+129     & 19.17 (0.38)&  & 0.57 (0.15) & -3.99 &                 1\\

1538+333     & 29.1 (1.1)& DA & 0.63 (0.03) & -3.06 &         \\
1540+236     & 19.6 (0.8)&  & 1.11 (0.1) & -4.2 &                                   1, 16 \\
1609+135     & 18.4 (1.6)& DA & 1.07 (0.06) & -3.5 &          \\
1626+368     & 15.9 (0.5)& DZA & 0.58 (0.03) & -3.13 &         \\
1630+089     & 13.8 (0.4)&  & 0.59 (0.15) & -3.81 &                     1, 9, 17\\
1632+177     & 18.7 (0.7)& DA & 0.46 (0.02) & -2.64 &         \\
1633+433     & 15.1 (0.7)&  DAZ& 0.68 (0.04) & -3.63 &         \\
1639+537     & 21.2 (1.6)&  & 0.62 (0.11) & -3.4 &                      9, 18\\
1647+591     & 10.98 (0.07)& DAV & 0.76 (0.03) & -2.55 &        3, 4, 5, 7, 8 \\
1653+385     & 30.7 (1.2)& DAZ & 0.59 (0.00) & -3.77 &         \\
1655+215     & 23.3 (1.7)& DA & 0.52 (0.06) & -2.9 &          \\
1657+321     & 51.7 (2.5)& DA & 0.59 (0.00) & -3.62 &         \\
1705+030     & 17.5 (1.7)& DZ & 0.68 (0.09) & -3.67 &         \\
1729+371     & 50.3 (2.2)& DAZB & 0.64 (0.03) & -2.8 &          \\
1748+708     & 6.07 (0.09)& DXP & 0.79 (0.01 & -4.07 &         \\
1756+143     & 20.5 (1.2)& DA & 0.58 (0.00) & -3.99 &         \\
1756+827     & 15.7 (0.7)& DA & 0.58 (0.04) & -3.39 &         \\
1814+134     & 14.2 (0.2)& DA & 0.68 (0.02) & -4.05 &         \\
1820+609     & 12.8 (0.7)& DA & 0.56 (0.05) & -4.06 &         \\
1829+547     & 15.0 (1.3)& DXP & 0.9 (0.07) & -3.94 &          \\
1900+705     & 13.0 (0.4)& DAP & 0.93 (0.02) & -2.88 &         \\
1912+143     & 35.0 (6.6)&  & 1.03 (0.09) & -3.89 &             1, 9\\ 
1917+386     & 10.51 (0.06)& DC & 0.75 (0.04) & -3.77 &  7, 8        \\
1919+145     & 19.8 (0.8)& DA & 0.74 (0.03) & -2.21 &         \\
1935+276     & 18.0 (0.9)& DAV & 0.6 (0.03) & -2.41 &   15      \\
1953$-$011   & 11.4 (0.4)& DAH & 0.73 (0.03) & -3.38 &         \\
2002$-$110   & 17.3 (0.2)& DC & 0.72 (0.01) & -4.29 &         \\
2007$-$303   & 15.4 (0.6)& DA & 0.6 (0.02) & -1.97 &          \\
2008$-$600   & 16.6 (0.2)& DC & 0.44 (0.01) & -3.97 &         \\
2032+248     & 14.6 (0.4)& DA & 0.64 (0.03) & -1.56 &       3, 4, 5\\
2039$-$202   & 21.1 (0.8)& DA & 0.61 (0.03) & -1.58 &         \\
2039$-$682   & 19.6 (0.9)& DA & 0.98 (0.03) & -2.27 &         \\
2040$-$392   & 22.6 (0.5)& DA & 0.61 (0.03) & -2.62 &         6\\
2047+372     & 17.3 (0.3)&  DA& 0.81 (0.03) & -2.34 &         \\
2048$-$250   & 28.2 (1.1)& DA & 0.59 (0.00) & -3.31 &         \\
2058+550 &      22.6    (2.5) &         &               0.58    (-)&     -& 2\\
\hline
\end{tabular} 
\label{tbl:obs_b}
\end{table*}

\begin{table*}
\renewcommand\thetable{2c}
\centering
\begin{tabular}{lccccc}
\hline
\hline
 & Distance [pc]& Spectral type & Mass [\Msolar] &  log L/\Lsolar & References \\
\hline
2105$-$820   & 17.1 (2.6)&  DAZH& 0.74 (0.13) & -2.93 &         \\
2115$-$560   & 26.5 (1.0)& DAZ & 0.58 (0.03) & -2.83 &         \\
2117+539     & 17.3 (0.2)& DA & 0.56 (0.03) & -2.1 &     7, 8     \\
2133$-$135     & 20.4 (3.5)&  & - &  - &                                        3, 6\\
2138$-$332   & 15.6 (0.3)& DZ & 0.7 (0.02) & -3.48 &                \\
2140+207     & 12.5 (0.5)&  DQ& 0.48 (0.04) & -3.09 &           \\
2159$-$754   & 21.0 (1.1)& DA & 0.92 (0.04) & -3.35 &           \\
2210+565     & 22.3 (1.4)&  & 0.68 (0.03) & -1.97 &             7, 18, 19\\
2211$-$392   & 18.7 (0.9)& DA & 0.8 (0.04) & -3.88 &                    \\
2215+368     & 23.5 (1.8)& DC & 0.58 (0.00) & -4.05 &           \\
2246+223     & 19.1 (1.5)& DA & 0.96 (0.06) & -3.13 &           \\
2251$-$070   & 8.520 (0.069)& DZ & 0.58 (0.03) & -4.45 &                3, 5, 6\\
2326+049     & 13.6 (0.8)&  DAZ& 0.63 (0.03) & -2.5 &                   \\
2336$-$079   & 15.9 (0.4)&  DAV& 0.76 (0.02) & -2.82 &          \\
2345+027 &      22.7    (3.6) &         &               0.58    (-)&    -& 2\\
2347+292     & 21.5 (1.9)& DA & 0.49 (0.08) & -3.69 &           \\
2359$-$434   & 8.169 (0.074)& DA & 0.78 (0.03) & -3.26 &            3, 5, 6\\
\hline
\end{tabular} 
\newline
\begin{flushleft} \tablefoot{ 
$^1$\citet{vanaltenabook95};
$^2$\citet{vanleeuwen07};
$^{3}$Discovery and Evaluation of Nearby Stellar Embers (DENSE) project, http://www.DenseProject.com;
$^{4}$\citet{subasavage09};
$^5$\citet{sion14};
$^6$\citet{gatewood09};
$^{7}$\citet{Tre17}; 
$^{8}$\citet{Gai16}; 
$^{9}$\citet{gianninas11};
$^{10}$\citet{subasavage08};
$^{11}$\citet{Tre11};
$^{12}$\citet{holberg13};
$^{13}$\citet{limoges13};
$^{14}$\citet{limoges15};
$^{15}$\citet{vanaltenabook95};
$^{16}$\citet{sayres12};
$^{17}$\citet{salim03};
$^{18}$\citet{gliese91};
$^{19}$\citet{holberg16}.
}\end{flushleft} 
\label{tbl:obs_c}
\end{table*}

\section{The effect of stellar wind on wide binaries}
\label{sec:app_wind}
\begin{figure}
    \centering
        \includegraphics[width=\columnwidth]{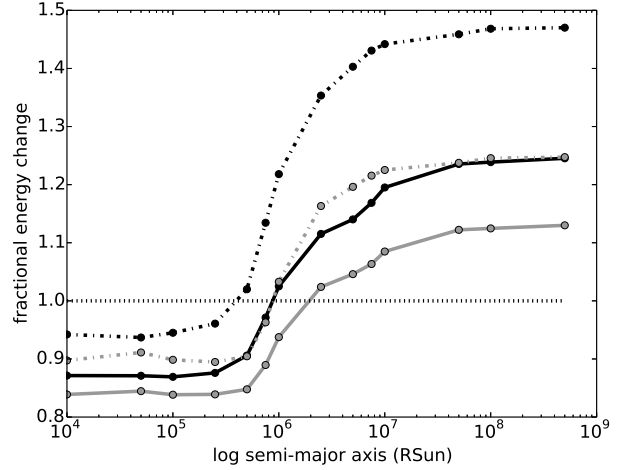} 
    \caption{Fractional energy change of the orbit due to winds as a function of initial orbital separation for initially circular orbits. The different lines represent four different systems. Low-mass ratios are shown in black, high-mass ratios in grey. The black, grey, black-dashed, and grey-dashed lines represent systems with initial masses of (2.5 \& 1\,\Msolar), (2.5 \& 2\,\Msolar), (5 \& 1\,\Msolar), and (5 \& 4\,\Msolar), respectively. If the fractional energy change is larger than unity, the system dissolves. }
    \label{fig:wind}
    \end{figure}

We simulate the dynamical and stellar evolution of wide binary stars using the Astrophysical Multi-purpose Simulation Environment (AMUSE). For the dynamical evolution we use a direct, fourth-order Hermite integrator \citep{Mak92}, and for the stellar evolution we use the same code as used for the BPS simulations in this paper (SeBa). Every integration time step, we evolve the dynamics and stellar evolution independently, after which we synchronize the data with the new updated masses, positions, and velocities. The time step criterion is based on changes in the masses of the stars, such that more steps are taken during events of rapid mass loss. The dynamical code has its own internal time step criterion to resolve close encounters, but will always finish on the prescribed integration time. We evolve the binary stars until the primary component has become a WD, and then we measure the final orbital energy of the system. 
If the fractional energy change $-(E_{\rm orb, final}- E_{\rm orb, init})/E_{\rm orb, init}$ exceeds unity, then the system dissolves. The four binary systems in Fig.\,\ref{fig:wind}, chosen to represent a wide range in WD binary progenitors, all dissolve if the initial separation is wide enough. The critical separation is of the order of $10^6\,\Ro$. For eccentric systems the outcome can be different and it is likely dependent on the orbital phase.

\end{appendix}

\bibliographystyle{aa}
\bibliography{bibtex_silvia_toonen.bib}

\end{document}